\documentclass{aa}
\usepackage[varg]{txfonts}

\usepackage{graphicx}

\def\nct#1{\nocite{#1}}

\def\psibe{\psi_B}
\def\ibm{I_{\rm mb}^\perp}
\def\ibmpar{I_{\rm mb}^\parallel}
\def\dbm{\delta_{\rm mb}}
\def\bm{\boldmath}

\def\aap{A\&A}

\def\apjs{ApJS}

\begin{document}

%%%%%%%%%%%%%%%%%

%\title[Distortions of polarisation in pulsars]
\title{A model for distortions of polarisation-angle curves\\ in radio pulsars}

\author{J.~Dyks,\inst{1}
M.~Serylak,\inst{2} 
S.~Os{\l}owski,\inst{3,4} L.~Saha,\inst{1} L.~Guillemot,\inst{5, 6}
%\newauthor 
I.~Cognard\inst{5, 6} and B.~Rudak\inst{1}
}

%\offprints{J. Dyks}

\institute{Nicolaus Copernicus Astronomical Center, 
Polish Academy of Sciences,
 Rabia\'nska 8, 87-100, Toru\'n,
Poland \and
University of the Western Cape, Cape Town, South Africa 
\and
Fakult\"{a}t f\"{u}r Physik, Universit\"{a}t Bielefeld, Postfach 100131,
33501 Bielefeld, Germany 
\and Max-Planck-Institut f\"{u}r Radioastronomie, Auf dem H\"{u}gel 69, 53121
Bonn, Germany 
\and
Laboratoire de Physique et Chimie de l'Environnement et de l'Espace
LPC2E CNRS-Universit\'{e} d'Orl\'{e}ans, F-45071 Orl\'{e}ans, France 
\and
Station de radioastronomie de Nan\c{c}ay, Observatoire de Paris,
CNRS/INSU F-18330 Nan\c{c}ay, France
}

\date{Received 2015 August 10; in original form 2015 July 23}

%\pagerange{\pageref{firstpage}--\pageref{lastpage}} \pubyear{2002}
%\maketitle
%\label{firstpage}

%\begin{abstract}
\abstract{Some radio pulsar profiles (in particular those of millisecond pulsars) 
contain wide emission structures which cover large intervals
of pulse phase. Local distortions 
of an average curve of polarisation angle (PA)
can be identified in such profiles, and they are often found to be associated
with absorption features or narrow emission components.}
{
 The features may be interpreted as a convolution of a lateral 
profile of an emitter with a 
microscopic radiation pattern of a non-negligible angular extent.}
{
We study a model which assumes that 
such an extended microbeam of the X-mode curvature radiation 
is spreading the radiation polarised at a fixed
position angle within an interval of pulse phase.}
{
The model is capable of 
interpreting the strongly dissimilar polarisation of double notches in 
 PSR B1821$-$24A (for which we present new polarisation data 
from the Nan\c{c}ay Radio Telescope) 
and PSR J0437$-$4715. It also explains a step-like change in PA
observed at the bifurcated trailing
component in the profile of J0437$-$4715. 
A generic form of the modelled PA distortion
is a zigzag-shaped wiggle, which 
in the presence of the second polarisation mode (O mode) 
can be magnified or 
transformed into a W- or U-shaped
deflection of a total net PA.}
{
The model's efficiency
in interpreting dissimilar polarisation effects provides further credence
to the stream-based (fan-beam) geometry of pulsar emission.
It also suggests that the microbeam width may not always 
be assumed negligible
in comparison with the angular scale of emissivity gradients in the
emission region.}
%\end{abstract}

\authorrunning{Dyks et al.}
\titlerunning{Model for polarisation distortions in radio pulsars}

%\begin{keywords}
\maketitle
%\end{keywords}

\def\lap{\hbox{\hspace{4.3mm}}
         \raise1.5pt \vbox{\moveleft9pt\hbox{$<$}}
         \lower1.5pt \vbox{\moveleft9pt\hbox{$\sim$ }}
         \hbox{\hskip 0.02mm}}

\def\rwobs{R_W}
\def\rwcon{R_W}
\def\rwstr{R_W}
\def\winobs{W_{\rm in}}
\def\woutobs{W_{\rm out}}
\def\phm{\phi_m}
\def\phmi{\phi_{m, i}}
\def\thm{\theta_m}
\def\dres{\Delta\phi_{\rm res}}
\def\win{W_{\rm in}}
\def\wout{W_{\rm out}}
\def\rin{\rho_{\rm in}}
\def\rout{\rho_{\rm out}}
\def\phin{\phi_{\rm in}}
\def\phout{\phi_{\rm out}}
\def\xin{x_{\rm in}}
\def\xout{x_{\rm out}}

\def\thmin{\theta_{\rm min}^{\thinspace m}}
\def\thmax{\theta_{\rm max}^{\thinspace m}}

\section{Introduction}
\label{intro}

Pulsar polarisation escapes thorough understanding despite
more than four decades of increasingly deep study,   both
observational and theoretical. The association of observed
polarisation angle (PA, hereafter  also denoted by  $\psi$)
with a projection of local magnetic field (Radhakrishnan \& Cooke 1969)
\nct{rc69}
%hereafter RC69)
has strengthened the magnetic pole model of a radio pulsar beam.
A mathematical formulation of this rotating vector model 
(RVM, Komesaroff 1970) \nct{k70} gives a chance
%makes it possible 
%\LEt{ ok? "chance" sounds very arbitrary } 
%{\bf [ok? "chance" sounds very arbitrary] No, please use the original wording.
%"gives a chance" is much closer to truth, because the 
%math works for a handful of pulsars out of 2.5 thousands.
%So the math usually does NOT make it possible. Because of the nature of
%data, the math does not work, except for few rare cases. 
%So just saying "makes it possible" would be misleading. 
%} 
to determine the global geometric parameters, such 
as the
magnetic dipole inclination $\alpha$ and the viewing angle $\zeta$,
both measured with respect to the rotation axis {\bm $\mit \Omega$}.

\begin{figure}
\includegraphics[width=0.48\textwidth]{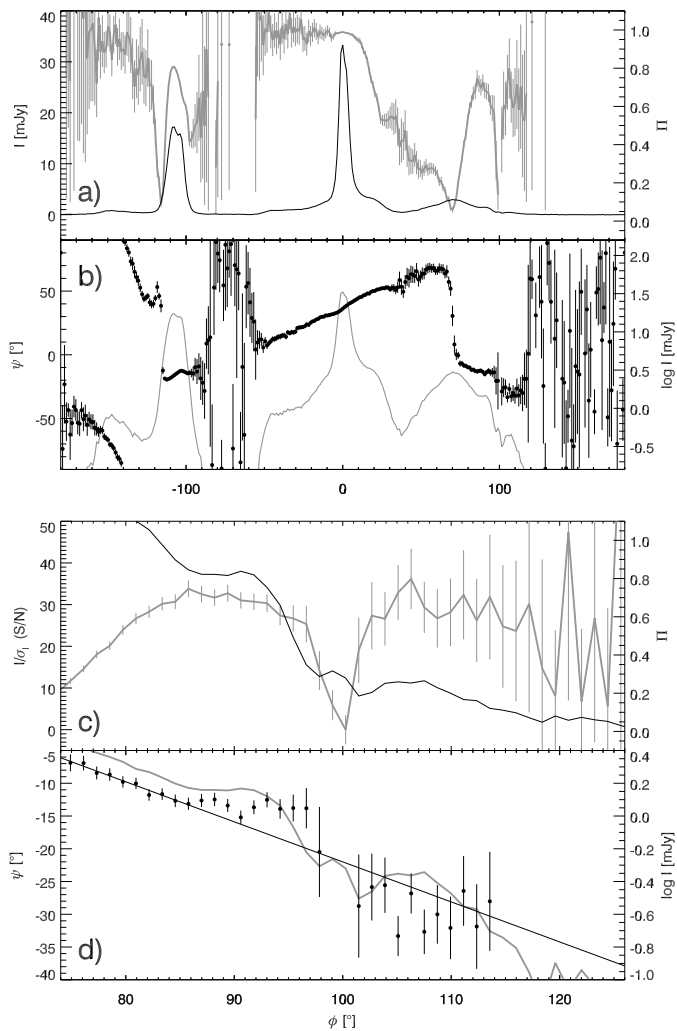}
\caption{Linear polarisation properties of PSR B1821$-$24A
as observed with GBT in L band (BPDR). Grey curves refer to the
vertical axes on the right-hand side. The bottom two panels zoom into
the double notches at $\phi\approx100^\circ$.
{\bf a)} Total flux $I$ (black solid) and the linear polarisation fraction
$\Pi=L/I$ (grey). {\bf b)} The PA (black) and $\log I$ (grey).
{\bf c)} The total flux S/N (black solid) and the polarisation fraction (grey).
{\bf d)} The PA (black) and $\log I$ (grey). The straight line presents
the PA variations anticipated in the absence of the notches. 
%We note 
%{\bf 
%Notice 
%(((Please do not use "We note".
%"We note" sounds very strange and puts unnecessary emphasis on "We".
%The meaning of "Note" was "Please notice", not "We note". 
%The feature described in the caption to Fig. 1 was also noticed by at least
%one other group of researchers, so "We note" would better not be used.
%I replaced "Note" with "Notice" everywhere.)))
%}
There is a drop in $\Pi$ and a change in PA at the notches.
The zero point of PA (whether observed or modelled) is arbitrary
in all figures of this paper.
}
\label{flux}
\end{figure}
\begin{figure}
\includegraphics[width=0.48\textwidth]{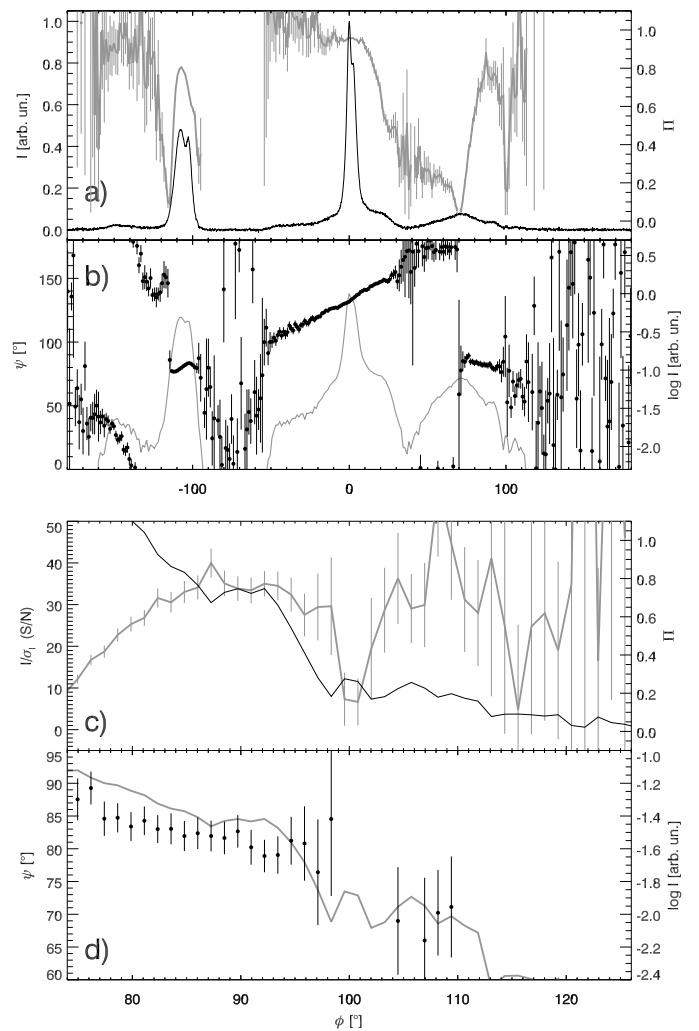}
\caption{Linear polarisation properties of PSR B1821$-$24A
observed at $1.5$ GHz with the Nan\c{c}ay telescope. 
The layout is the same as in Fig.~\ref{flux}.
The total profile in a) has 2048 samples per period $P$, whereas 
 resolution of the other 
data was decreased to 292 bins/$P$ by merging seven adjacent bins.
%are for seven adjacent bins merged (292 bins/$P$). 
%\LEt{ seven merged bins, seven bins that are merged? the other data are merged for the seven adjacent bins? }
%{\bf [seven merged bins, seven bins that are merged? the other data are merged 
%for the seven adjacent bins?] 
%I tried to rephrase as above.
%We merged bins: 1 to 7, 8 to 14, 15 to 21, etc. (up to the bin 2048).
%"Seven merged bins" is not the best wording, because all 
%bins in the original vector (2048 of them) went through the merging
%procedure, not just seven. Therefore, merging of "seven adjacent bins"
%seems to be better than merging of "seven bins".
%}
}
\label{lucas}
\end{figure}

Because of the commonly encountered deviations of the observed PA from the
simple RVM model, several extensions of the model have been made.
Blaskiewicz et al.~(1991) 
\nct{bcw91}
 included the special relativistic
effects on the PA curve (see also Dyks 2008), \nct{d08}
whereas Hibschman and Arons (2001) \nct{ha01} also included
the modification of a local magnetic field by
magnetospheric currents.
Statistical studies of single-pulse PA distributions
(McKinnon \& Stinebring 1998, van Straten 2009) 
\nct{ms98, v09}
have shown the
importance of various ways in which two polarisation modes 
can be combined in the presence of observational noise. 
Single-pulse analysis of polarisation (Edwards, Stappers \& van Leeuwen
2003; Rankin \& Ramachandran 2003) 
\nct{esv03, rr03}
has shown a spatial (angular) or temporal separation of orthogonal
polarisation modes, interpreted either in terms of different
refraction properties of these modes (Petrova \& Lyubarski 2000;
 Lyubarsky 2008) \nct{pl00, l08} or
different locations of the modes in the radiation pattern of a specific
emission mechanism (Dyks et al.~2010, hereafter DRD10). \nct{drd10}     
On the theoretical side, the convolution of 
a spatially extended emission with microphysical 
radiation beams has been numerically
studied for non-coherent emission processes 
(Wang et al.~2012, Kumar \& Gangadhara 2012) \nct{wwh12, kg12}
with the inclusion of possible propagation effects on the observed
polarisation properties (Barnard \& Arons 1986; Wang et al.~2010; 
Beskin \& Philippov 2012).
\nct{ba86, wlh10, bp12}
In these studies, the angular size of the microphysical radiation
pattern was negligible in comparison to the angular gradients of emissivity
in the emission region. Instead, in this paper we assume that  
these scales are comparable. 

Via statistical modelling of single-pulse effects, 
Luo (2004) 
\nct{luo2004}
has shown that the microbeam effects are capable of generating 
perceivable non-RVM distortions when radio waves decouple from 
the local plasma close to the emission region. 
Melrose et al.~(2006) 
\nct{mmk2006}
have shown that statistical properties 
of single-pulse emission in both polarisation modes 
considerably influence the PA distributions observed at a fixed phase,  
hence they affect the average PA curve. By assuming appropriate populations
of single pulses, they were able to reproduce non-trivial
distribution of data on the Poincare sphere.

In spite of these developments, it is usually impossible to explain
why specific deviations of the average PA curve from the RVM model 
are observed.
Notable exceptions are the orthogonal-mode jumps in PA by 
about $90^\circ$, which clearly originate from one mode being
overtaken by its orthogonal counterpart (Cheng \& Ruderman 1979, 
hereafter CR79; Tinbergen 2005). \nct{cr79, t05}
In this paper 
we attempt to understand less obvious distortions of PA,
using a simple physical model of a polarised radiation beam,
and convolving it with a macroscopic 
spatial distribution of emissivity.  
Our approach may be considered complimentary to that of Melrose 
et al.~(2006) who modelled the statistical effects of polarised 
single pulses at a fixed phase. 
We instead focus on how the microbeam topology 
relates the PA at adjacent pulse longitudes while ignoring most of single-pulse effects. The non-RVM PA distortions may also be caused 
by multiple (Mitra \& Li 2004) 
\nct{ml2004}
or radially extended (Dyks 2008) \nct{d08}
emission heights, return currents (Ramachandran \& Kramer 2003),
\nct{rk2003}
and scattering in
the interstellar medium (Karastergiou 2009). \nct{k2009}
None of these effects is 
 included in the present study. 
Our model has been inspired by the high-quality observations
of double notches reported 
for PSR B1821$-$24A in Bilous et al.~(2015, hereafter BPDR;
see Figs.~1 and 2 therein). \nct{bpd15} 
However, the model is also applicable to the notches 
and the bifurcated emission component observed on the trailing side of the
profile of PSR J0437$-$4715 (Navarro et al.~1997). \nct{nms97}

Section \ref{obser} describes available polarisation data 
for pulsars with double features in their average profiles. 
It is then followed by the description of model assumptions 
and the numerical method (Section \ref{model}).
Section 4 presents the results of the model simulations, which 
%{\bf 
% are 
%(((Seems the word "are" got lost)))
%}
  are immediately compared to the observations.

\section{Observed polarisation of profiles with double features}
\label{obser}

Bilous et al.~(2015) \nct{bpd15} 
%BPDR 
%\LEt{ avoid abbreviations at the beginning of  a sentence. Spell out here and check throughout. } 
reported high S/N L-band polarimetric observations of 
a 3.05-millisecond pulsar B1821$-$24A, 
which are summarised in our Fig.~\ref{flux}.
The profile mostly exhibits a high linear polarisation fraction $\Pi=L/I$
except for two quasi-orthogonal polarisation jumps 
at $\phi\approx-117^\circ$ 
and $72^\circ$. A considerable drop in $\Pi$, however, also occurs
at $\phi\approx 100^\circ$ where the double notches with a low, shallow
central bump %maximum 
are found. Interestingly, even though $\Pi$ decreases
 to nearly zero (grey line in Fig.~\ref{flux}c)
the PA changes just a little bit, by approximately $15^\circ$ (Fig.~\ref{flux}b and
d). Although this change in PA has been described as a jump in BPDR,
it may also be interpreted as a zigzag-shaped wiggle around
a monotonically decreasing PA (marked with the straight line in
Fig.~\ref{flux}d). 
This work provides a framework which supports this interpretation.
%This will be the interpretation developed in the modelling
%part of this paper. 
%On the right-hand side of the notches, at
%$\phi \ga 115^\circ$, the PA steeply increases, probably
%entering another orthogonal PA jump at $\phi\approx120^\circ$.
%Data quality becomes very low there, so we exclude the region with $\phi >
%115^\circ$ from analysis.
In addition to the PA change at the phase of $100^\circ$, the profile of
B1821$-$24A exhibits other interesting deflections from a smooth PA curve.
There is a zigzag-like wiggle at the second-brightest 
component (P1, located at $\phi = -107^\circ$)
and a one-directional PA deflection 
near the brightest component (P2, located at 
$\phi = 0^\circ$). 
They are associated with a larger $\Pi$ 
and deviate from a smoothly interpolated PA curve 
%by a smaller amount of PA than 
less than 
observed at the double notches. Both P1 and P2 reveal a double structure,
although they are not as well resolved and not as symmetric 
as the prominent double features observed in other
pulsars (e.g.~the bifurcated precursor of J1012$+$5307, DRD10).
The slight PA distortions associated with these emission components
will be addressed in Section \ref{emicom}.

%\subsection{Nan\c{c}ay observations of B1821$-$24A}

\subsection{Observations of B1821$-$24A with the Nan\c{c}ay Radio Telescope}

The S/N of the profile within the notches observed by BPDR is $\approx 10$
(black solid line in Fig.~\ref{flux}c), which is relatively
low. This persuaded us to confirm the presence of the notches
%\LEt{ the presence of what? the notches? } 
 using independent
observations. PSR B1821$-$24A is regularly observed with the 100 m
equivalent Nan\c{c}ay Radio Telescope (NRT) as a part of its pulsar
timing program. In order to obtain high S/N profile we  used 82
observations 
taken between August 2011 and May 2015 approximately once every two weeks.

The  data were recorded with the NUPPI backend (Liu et al.~2014) \nct{ldc14}
%(Desvignes et al. 2013) \nct{dcc13}
-- a flexible digital signal processor designed for pulsar observations.
First, Nyquist sampled spectra were acquired centred at 1484 MHz in 512 MHz
bandwidth
with 4 MHz resolution producing 128 frequency channels. The data were
coherently
dedispersed 
%\LEt{ re? redispersed? } 
%{\bf [re? redispersed?] No, "dedispersed" is the correct word 
%(a standard word for the procedure of 
%removal of radio waves' dispersion).
%} 
in real time using a dispersion measure value of 119.894 pc
cm$^{-3}$ in
order to correct for the dispersive delay caused by the interstellar medium
(ISM).
The data were then transformed from XY auto- and cross-correlations to full
Stokes parameters, folded into final full-Stokes pulse profiles with
resolution of
2048 samples per period and written  out to disk in PSRFITS (Hotan et al.
2004)  \nct{hvm04} 
format. Each of the 82 observations was then inspected offline for the
presence of
radio-frequency interference (RFI) using standard pulsar processing tools
({\sc PAZ}, {\sc PSRZAP}) from the {\sc PSRCHIVE} pulsar processing suite
(Hotan et al. 2004) by zero-weighting the affected portions of the data.

Flux and polarisation calibration were also performed  using the PSRCHIVE
software package.
A locally generated standard pulsed noise source was recorded to determine
the flux 
%\LEt{ flux?  also next sentence } 
scale.
The equivalent noise source flux density was obtained by observation of the
unpolarised 
quasar 3C48. Regular observations of linearly polarised pulsar B1929+10 were
used to
characterise the receiver system's intrinsic polarisation cross-coupling
matrix
(van Straten 2004). 
\nct{v04}
The data was then corrected for the effects of rotation
measure
(RM) using the PSRCHIVE tool {\sc RMFIT}. This resulted in an average profile
with
a total integration time of 49.2 hours ($\approx 5.8 \cdot 10^{7}$ periods).

The black solid line in panel a of Fig.~\ref{lucas} 
presents the average profile with
2048 samples per period. 
To increase the S/N, the rest of the observables in Fig.~\ref{lucas}c 
are plotted after merging seven adjacent bins 
%with seven adjacent bins merged, 
%\LEt{ see note 3 } 
%{\bf  [see note 3] Rephrased.} 
for a total of 292 samples per period, which is 
close to the resolution of the BPDR profile.
% (298 samples per $P$).
Although the S/N of the Nan\c{c}ay profile is slightly lower,
it does confirm the existence of the double 
%{\bf 
`absorption'
%(((Please use the apostrophes here and in one more place two paragraphs
%below; in the other places the word absorption was used in the normal sense, 
%and these should be without apostrophes.
%It is not known if the notches are an absorption feature 
%or a feature that results from a lack of
%emission. This is why apostrophes were used, so that the meaning 
%of 'absorption' (in apostrophes) is: "absorption of emitted radiation 
%or the lack of emission without any absorption involved". 
%I am not aware of any single word that reflects both these possibilities 
%and would be properly understood by pulsarists, and the apostrophes were 
%successfully used before. The use of 'absorption' is also justified
%by the similarity of these features to absorption lines in star spectra.)))
%}
 feature 
at $\phi\approx 100^\circ$. As in BPDR, the two minima seem to be separated by an
indistinct bump.  %maximum. 
Our data also confirm the change in PA, and the 
steep decrease in $\Pi$ nearly  to zero at the centre of the notches.
 In the Nan\c{c}ay profile  P1 and P2 are both clearly double. 
%\LEt{ ok? to avoid the symbol at the beginning of the sentence }

Polarised profiles of B1821$-$24A are also published by
Dai et al.~(2015),
 \nct{dhm15}
who present the Parkes data at $10$, $20$, and $50$ cm.
The change in PA at the phase of the notches %which lags P2 by $100^\circ$
is detectable there; however, the double form of the notches or of P1 and P2
is not resolved. 
The linear polarisation fraction $\Pi$ considerably decreases
within the notches (from $0.72$  to $0.37$, Fig.~A16 therein), 
although not down to the near-zero value reported in BPDR.
The difference may result from the smearing 
apparent in the Parkes profile.\footnote{The minimum $\Pi$ measured by BPDR 
is eight times smaller than its
statistical error, hence this data point is not plotted in panel a
of Fig.~\ref{flux}, which only shows the points with $\Pi > \sigma_\Pi$, 
where $\sigma_\Pi$ is the statistical error of $\Pi$.}
%\LEt{ a full sentence should not be used with square brackets. If you feel this sentence should not be in the text you can put it into a footnote (but remove "We note that" because a footnote is already a note)  }
%Had it been used consistently,
%the dotted part of $\Pi$ in Fig.~\ref{flux}c should not be plotted.

Generally, all available data are consistent with the existence of
double 
%{\bf 
`absorption' 
%} 
feature at $\phi\approx100^\circ$, accompanied by the
considerable drop in $\Pi$ and a modest change in  the PA value.
Bilous et al.~(2015) note that giant pulses 
in B1821$-$24A occur at the phase of the
double notches, and a peak of a narrow component in an X-ray profile 
of this pulsar nearly coincides with the notches.

\subsection{Linear polarisation of PSR J0437$-$4715}

\begin{figure}
\includegraphics[width=0.48\textwidth]{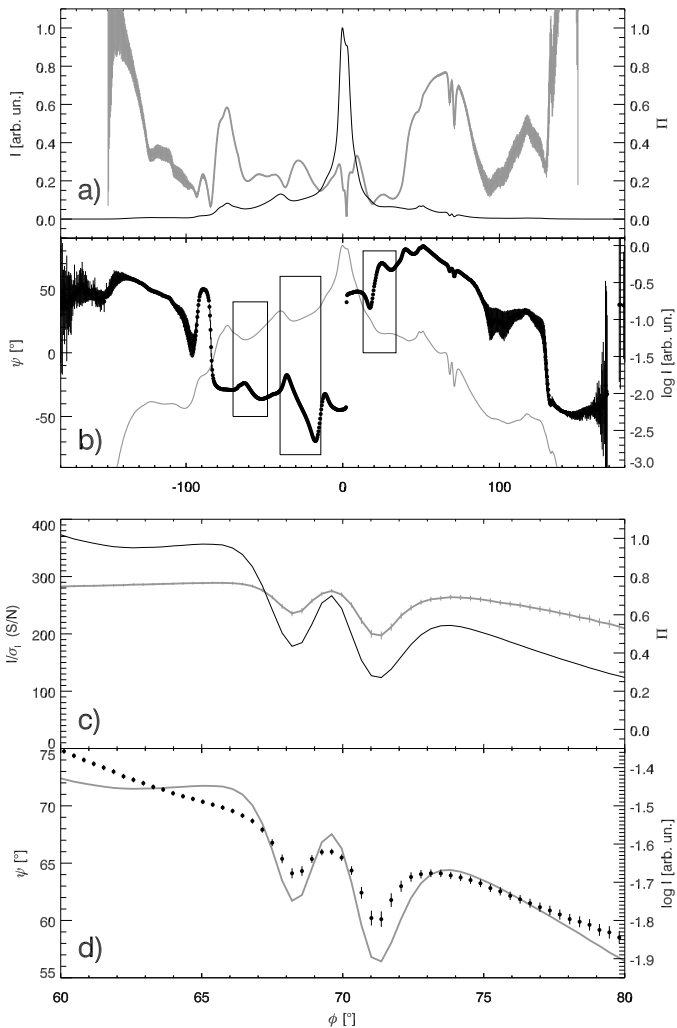}
\caption{Linear polarisation properties of PSR J0437$-$4715
observed at 1.4 GHz with the Parkes telescope (Dai et al.~2015). 
The layout is the same as in the previous figures. The rectangles in b) mark 
the zigzag-shaped PA distortions that are discussed in Section \ref{monstersec}.
}
\label{monster}
\end{figure}
\nct{dhm15}

Another pulsar that will be discussed in terms of our model
 is PSR J0437$-$4715, a 5.25 ms pulsar with a long observation record at the
Parkes telescope (e.g.~Navarro et al.~1997; Oslowski et al.~2014; 
Dai et al.~2015). \nct{nms97, ovb14, dhm15}
Its polarisation properties
 exhibit interesting similarities to and differences from the case of
B1821$-$24A. Figure~\ref{monster} summarises the best available linear 
polarisation data from Dai et 
al.~(2015).\footnote{The data were obtained from the Parkes
 Observatory Pulsar Data Archive \nct{hmm11} (Hobbs et al.~2011,  
dx.doi.org/10.4225/08/54F3990BDF3F1). 
}
\nct{dhm15}
Double notches in J0437$-$4715 lag behind the peak of its brightest component
 by approximately $69^\circ$
and have a pronounced central maximum, reaching a large fraction
of the flux observed outside the notches.
The linear polarisation fraction $\Pi$ is decreasing at the minima of the
notches by 20-30 per cent 
%{\bf 
of
%(((Is this correct? For last 20 years language editors were always removing
%this "of" from my papers (to get the form: "X per cent something"). 
%Now, when I stopped using it, it has been inserted.)))
%}
 the off-notch value (from 0.75  to 0.6 
at the leading-side notch, and by more than a quarter at the trailing notch; 
see the grey line in Fig.~\ref{monster}c). 
At each minimum, the PA deflects toward the same
direction by $\Delta\psi \approx 5^\circ$,
which is a different behaviour from B1821$-$24A.
Several other deflections of PA are observed across the profile,
including the zigzags in the phase intervals marked with the rectangles in 
Fig.~\ref{monster}b:
 $\phi \in (-70^\circ,-48^\circ)$, 
$\phi \in (-40^\circ,-10^\circ),$
and $\phi\in(10^\circ,32^\circ)$. These wiggles seem to coincide in phase 
with broad minima in total intensity, centred at $\phi \approx -60^\circ$, 
$-25^\circ$, and $20^\circ$.

There are two more pulsars which exhibit both double notches
and PA deflections: B1929$+$10 %(Rankin \& Rathnasree 1997) 
and B0950$+$08. %(McLaughlin \& Rankin 2004).
In the first, the notches are observed at a flux level hundreds of
times lower than the peak of its main pulse. Well-calibrated polarisation data
for this weak emission component are not available ($L > I$ in
Rankin \& Rathnasree 1997). \nct{rr97}
In the case of B0950$+$08 the polarisation at $1.4$ GHz 
was published in McLaughlin \&
Rankin (2004), \nct{mr04} 
however, for a profile with almost invisible notches,
and with a small S/N in $L$.

\section{A model for the polarisation angle deflections}
\label{model}

%\textcolor{blue}{ 
%}

%\def\delb{\delta_{bm}}

\begin{figure}
\includegraphics[width=0.48\textwidth]{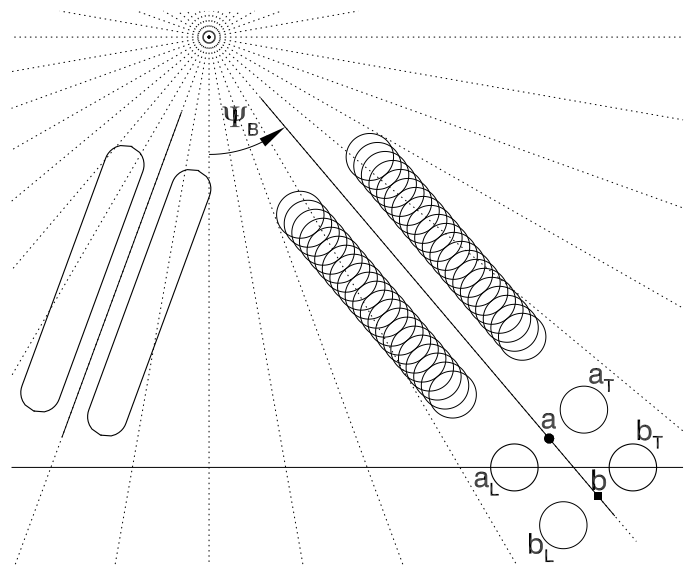}
\caption{Sky-projected 
%head-on
view of the split-fan beams 
typical of the X-mode curvature radiation from narrow plasma streams. 
The continuous beam shown on the left can be decomposed into
a sequence of lobe pairs, emitted quasi-instantaneously from different
points along the stream (shown on the right-hand side), 
e.g.~from the points `a' and `b' in the bottom right
corner.
The horizontal line marks the passage of the line of sight. 
The stream's polarisation angle $\Psi_B$
is detectable at two pulse longitudes (corresponding to the 
lobes a$_{\rm L}$ and b$_{\rm T}$) 
on the leading and trailing side of the stream.
}
\label{toon}
\end{figure}

The model assumes that the generic element 
of a radio emission region in a pulsar magnetosphere
has the form of a plasma stream occupying a narrow magnetic flux tube
(see Fig.~1 in Dyks \& Rudak 2012, hereafter
DR12). \nct{dr12} In a short time interval  
%($\Delta t \sim c \rho_{\rm crv}/\gamma$, where $...) 
(quasi-instantaneously), the 
charges in the stream emit a narrow pattern of
radiation beamed nearly along the velocity vector and pointed tangentially
to the stream. The angular size of such a microbeam is small but significant, 
and it needs
to be convolved with the spatial extent of the emission region 
(e.g.~with the lateral profile of density or emissivity in the stream).
When modelling the polarisation of notches 
we assume that the emitted microbeam
has the form of two lobes emitted at a small angle with respect to the
plane of the magnetic field (see Fig.~1 in DR12, 
and Figs.~10 and 11 in
DRD10). No radiation is emitted within the plane of the electron trajectory
(which in this paper is assumed to coincide with a $B$-field line).
Therefore, charges sweeping along a bent magnetic field, emit
a split-fan beam shown on the left-hand side of Fig.~\ref{toon}.
%Note that the angular separation $2\dbm$ between the sky-projected 
%emission lobes
%does not increase with the distance from the dipole axis, since it is
%determined by the emission physics (specifically by the local radius 
%of curvature of $B$-field lines, $\rho_{\rm crv}$, and the observed 
%frequency $\nu$). 

Because  the beam is double, each point of the stream
(e.g.  point `a' in Fig.~\ref{toon}) creates two bright patches of emission
on the sky (a$_{\rm L}$ and a$_{\rm T}$). Therefore,
a fairly localised
piece of emission region, (e.g.~the stream segment between 
`a' and `b' in Fig.~\ref{toon}) can be detected at two 
different phases in the profile. For the sightline passing
along the horizontal path in Fig.~\ref{toon}, the a-b segment will be
detected at the lobe a$_{\rm L}$ and b$_{\rm T}$.

The process of curvature radiation (CR) in the extraordinary polarisation
mode (hereafter the X mode or $\perp$-mode) 
provides an example of such a double-lobed beam.
%\footnote{In the case of CR, it holds that 
%$\dbm \propto \rho_{\rm crv}^{-1/3}$, 
%so with the increasing altitude
%(and the distance from the dipole axis) the lobes in the split-fan beam
%should actually aproach each other (not shown in Fig.~\ref{toon}).} 
 The CR beam emitted in vacuum
can be mathematically decomposed into two parts: a filled-in pencil
beam polarised in the plane of the $B$-field and the bifurcated
beam polarised at the right angle to the plane of $B$ 
(see Eq.~6.29 in Rybicki \& Lightman 1979; Konopinski
1981, p.~305). \nct{rl79, k81} 
In a strongly magnetised plasma,
the emitted beam gets decomposed into a similar filled-in part 
polarised parallel to the $B$-field line plane (O mode)
and a bifurcated X-mode part. In the limit of plasma 
in infinite magnetic field, only the X mode is emitted, again with 
the bifurcated beam pattern, and polarised at a right angle with respect to
the $B$ plane (Gil et al.~2004).
\nct{glm04}
To study the qualitative implications for the observed PA, 
it is sufficient 
to use the vacuum formula for the beam shape, which we do in
Sect.~\ref{numet} (Eq.~\ref{iper}).

% \nct{rl79, k81} 
%(Konopinski 1981).

%For the polarisation of rays in the model microbeam 
%we assume the properties of the CR.
% The net polarisation of the effective X-mode beam, 
%i.e.~the polarisation 
%recorded after the beam has fully swept across our line of sight,
%is orthogonal to the plane of the $B$-field line (Lyubarsky 2008; 
%Gil et al.~2004). 
%\nct{l08, glm04}
%We emphasize that this property (polarisation direction), 
%as well as the bifurcated form of the beam, 
%are common for both the CR in magnetised plasma, 
%as well as for the orthogonally-polarised part of the vaccum CR. 
%Therefore, 

In the quasi-instantaneous X-mode beam, however, the radiation
is polarised in the {\bm $\mit k \times B$} direction (see Fig.~1 in CR79
%Cheng \& Ruderman 1979, 
\nct{cr79}
or Fig.~16 in DRD10), where {\bm $\mit k$} is the wave vector
pointing towards the observer. The quasi-instantaneous 
polarisation direction then 
strongly depends on where exactly our line of sight is piercing the
microbeam.   
A moving charge is passing this
microbeam through our line of sight in a nanosecond timescale 
($\Delta t \sim 1/\nu$, where
$\nu$ is the observed frequency). During that time,
a limited segment of the charge's trajectory of length $\Delta s \sim 
c \gamma^2 \Delta t$ is exposed to the observer. For the charge's 
Lorentz factor of $\gamma \sim 10^2$, the detectable segment is few kilometers in length. The horizontal motion of the sightline
through the lobe a$_{\rm L}$ in Fig.~\ref{toon} is several orders of
magnitude slower than the fast sweep of the lobe along 
the guiding magnetic field line. Therefore,  
%\emph{
if the emissivity does not change along that detectable part 
of electron trajectory, 
%}
%\LEt{ no italics for emphasis. Please remove this formatting from your paper } 
%{\bf [no italics for emphasis. Please remove this formatting from your
% paper] I have removed all "font emphasis". Did I use wrong fonts,
%or no font-emphasis is possible in AandA?
%}
the observer is exposed to a time-symmetric signal of intensity
 and an antisymmetric signal of PA. 
%The latter has a shape similar to
%the RVM swing, albeit it has the microscopic origin and lasts for a
%nanosecond. 
These symmetries ensure that the net PA is orthogonal
to the projected $B$-field for an arbitrary location of sightline 
in the fully swept-by beam (the split-fan beam).\footnote{This condition 
of uniform emissivity applies to each
electron individually. 
%Different electrons, or different bunches of
%electrons, located in different parts of the stream, 
%may have different emissivity. The longitudinal profile of
%macroscopic emissivity in the stream, which is affected, 
%for example, by the local electron density, may be arbitrary. 
%Therefore, 
However, the single-pulse emission, as determined by the
longitudinal plasma density profile in the stream among others, may exhibit arbitrary
variability.
For this reason, the model is consistent with the observed randomness 
of subpulse
shapes (cf.~Sect.~3.2 in Luo 2004). 
However, it is not generally 
applicable for any pulse components, in particular those for which
the single-charge emissivity is variable on the timescale 
of $\gamma^2 \Delta t$ or those for which our sightline is just grazing
the periferies of the full split-fan beam.}

%However, when the electron's emissivity changes noticeably 
%on the angular scale comparable to the microbeam size, the symmetry is lost
%and  
%the microscopic {\bm $\mit k \times B$} direction of polarisation does not average
%out. It will be detectable as PA variations which are likely faster than
%those implied by the RVM.\footnote{These are not visible in the simulations
%of Wang et al.~(2015), because their microbeam is much narrower 
%than the angular scale of intensity gradients in the beams they model.
%See section 2.2 in DR12
% for a discussion of the microbeam scale problem.} 
%Therefore, a strict and general analysis requires
%a full 3D convolution of microbeams. 
% Such a fast change of emissivity is normally ignored for single electrons, 
%and it will not be considered in this paper, either.
%However, it is conceivable for quickly diluting  bunches 
%of charges.

%Following the well-established properties of CR in magnetised plasma, 
%in this paper we assume
%that the polarisation direction at any place in the \emph{effective} 
%microbeam, i.e.~the one which has fully passed through our line of sight, 
%is orthogonal to the sky projected guiding magnetic field 
%(i.e.~the field along which the emitting
%electron is moving). 
%The present version of the model 
%is then expected to work for pulse components 
% dominated by emission of the above-described type: 
%within the detectable 
%part of their trajectory, individual electrons
%need to emit at a steady rate. 

Accordingly, the lobe a$_{\rm L}$, 
 after fully passing through our line of sight, 
%emitted from the point `a' in Fig.~\ref{toon}, 
will contribute the PA fixed at the phase-independent 
value of $\psibe=\Psi_B+90^\circ$, 
where $\Psi_B$ is the PA corresponding 
to the sky-projected direction of the magnetic field at  point `a'.
Analogically, the trailing-side lobe b$_{\rm T}$ will be polarised at the same
angle $\psibe$, because the same magnetic azimuth corresponds to 
the emission point `b'. If the emissivity and curvature of the $B$-field line
do not change considerably between  points `a' and `b',
a symmetrical double-peaked emission component will be observed.
It will be highly polarised and have a phase-independent PA. 
%as long as we stay limited to the single (extraordinary) 
%polarisation mode.

The key assumptions of our model are then the following:
1) At least some parts of 
%the 
%{\bf (((I removed "the", because I mean any average profiles here, not
%necessarily those illustrated in Section 2 or discussed above.)))
%}
average profiles (especially those which exhibit 
bifurcated
features) consist of elementary radiation patterns 
which have the double-peaked cross section such as shown by the
dot-dashed line in Fig.~\ref{polidea}a. This one-dimensional version
of the microbeam pattern (denoted $\ibm$)
needs to be convolved with the macroscopic distribution of emissivity
$\eta_\perp$
within the radio-emitting region;
2) It is assumed that both lobes of the microbeam are highly polarised
at a fixed angle (of $90^\circ$ for the X mode) with respect to the
projected direction of the magnetic field line 
%\emph{
at the emission point
corresponding to the observed lobes.
%} 
A reference will be provided by the RVM PA value 
for the X mode
($\psi_B$) or the O mode ($\Psi_B$).

\subsection{Numerical method}
\label{numet}
 
Double notches and bifurcated emission components
are believed to be caused by a single void or peak
in the spatial emissivity profile $\eta_\perp$.
The double form
%\LEt{ The double/dual nature? Doubleness is not usually a word in English. If standard in your field, please ignore the suggestion } 
results from the bifurcated nature of the microbeam
which is convolved with this spatial emissivity
(DRD10; DR12; cf.~older models
based on different concepts: Wright 2004; Dyks et al.~2007).
\nct{w04, drr07}
Therefore, we start with a choice of the macroscopic emissivity $\eta_\perp$ 
as a function of phase.
In most cases we assume that $\eta_\perp$ has 
a narrow Gaussian dip (or peak) carved in (or projecting from) a uniform
emission component. An example of this $\eta_\perp$ 
is presented by the thick solid line 
in Fig.~\ref{polidea}a. 
This emissivity is convolved with the perfectly symmetric elementary
emission pattern (dot-dashed line in Fig.~\ref{polidea}a).
We neglect a possible asymmetry of this micropattern,
that may appear while the charges move between points `a' and `b' in
Fig.~\ref{toon}. Such asymmetry (and a related distortion of PA) 
is generally expected
when the radio emissivity varies with altitude and the sightline traverses
through the split fan at an oblique angle (see Fig.~2 in DR12). 
Therefore, the dot-dashed line in Fig.~\ref{polidea}a 
presents an effective pattern of the microbeam, 
mostly formed by emission from different points (`a' and `b' in
Fig.~\ref{toon}). 
 For the effective microbeam shape we take that
part of the vacuum
CR beam which is polarised orthogonally to the plane of a $B$-field line
\begin{equation}
\ibm \propto \xi \ K^2_{1/3} (y) \ \sin^2{(\phi-\phi^\prime)},
\label{iper}
\end{equation}
where
\begin{equation}
\xi=1/\gamma^2 + (\phi-\phi^\prime)^2, \ \ \ \ \ \ \ \ \  
y=\frac{2\pi\nu\rho_{\rm crv}}{3c}\ \xi^{3/2}, 
%\label{iper}
\end{equation}
$K$ is the modified Bessel function, $\phi^\prime$ is the phase of the microbeam centre,
and $\gamma$ is the Lorentz factor of the emitting particles 
(Rybicki \& Lightman 1979).
\nct{rl79} 
%This beam has the same qualitative properties (bifurcation, lack of 
%emission in the plane of charge trajectory, and
%polarisation direction) as the X-mode CR beam in magnetised plasma. 
%In particular, the shape 
%given by eq.~(\ref{iper}), and the angular separation of emission lobes, 
%are
%similar to the one calculated for the case of plasma in infinitely strong
%magnetic field (Gil et al.~2004). 
The magnification of the apparent microbeam through the non-orthogonal 
sightline cut at a possibly small $\alpha$ (section 2.2 in DR12) is ignored,
 i.e.~the observed width of the microbeam $\Delta$ 
is set only by the values of $\rho_{\rm crv}$ 
and $\nu$. 
The latter are selected to make the microbeam a few degrees wide
in the observed profile. 
The Lorentz factor $\gamma$ is fixed at a large 
(but otherwise arbitrary) value,
because neither the scale nor shape of the microbeam
depend on $\gamma$, whenever $\gamma \gg (\nu\rho_{\rm crv}/c)^{1/3}$. 
To perform the convolution, %of the Stokes parameters $I$, $Q$, and $U$,
a prescription for the $B$-field-based (RVM) PA curve is selected.
Since we focus on narrow phase intervals, we %usually 
approximate the PA
with a linear function of phase (straight dotted line in
Fig.~\ref{polidea}b).
For each pulse phase $\phi$, the contributed flux 
of the microbeam (centred at an arbitrary phase $\phi^\prime$)
is scaled by $\eta_\perp(\phi^\prime)$ and it is assumed to be 
fully linearly polarised
at the angle $\psibe(\phi^\prime)$. 
Thus, %no radiation from the microbeam centered at $\phi$ 
%is contributed at $\phi$. 
most of the flux cumulated at $\phi$
originates from two nearby phases on both sides of $\phi$.
Since these phases contribute their own values of $\psibe$,
the intrinsic distribution of PA at a fixed $\phi$ 
has a double-peaked form 
(Fig.~\ref{polidea}b), with no radiation
corresponding to the central value of $\psibe(\phi)$.
However, the average value of the observed PA, (denoted by $\psi_\perp$
since it refers to the X mode only), is equal to $\psibe(\phi)$.
When doing the convolution, we make the normal transitions between
$I$, $L=(Q^2+U^2)^{1/2}$, 
%{\bf 
%(((Please replace "tan" with "arctan" in the following formula for psi:)))}
$\psi=0.5\arctan(U/Q)$
and the Stokes parameters $I$, $Q=L\cos(2\psi)$, $U=L\sin(2\psi)$.
The circular polarisation and propagation 
effects are not studied in this work.

\section{Results}
\label{results}

In this section we present typical PA distortions 
implied by the model and discuss them in the light of observations
described in Sect.~\ref{obser}. 
When we refer to data, model parameters are
selected manually to obtain a qualitative agreement.
 
\subsection{Origin of PA deflections at double notches} 

\begin{figure}
\includegraphics[width=0.48\textwidth]{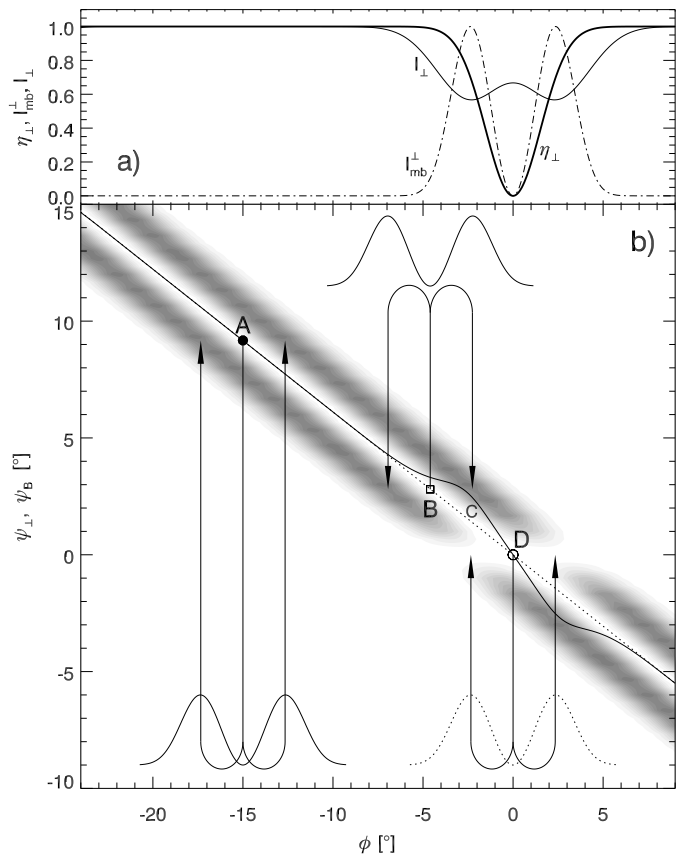}
\caption{Mechanism of the bidirectional (zigzag-shaped)
PA distortion. {\bf a)}  Effective microbeam pattern $\ibm$ (dot-dashed line),
and the macroscopic X-mode emissivity $\eta_\perp$ (thick solid) are
convolved into the net intensity profile $I_\perp$ (thin solid).
{\bf b)}  PA as a function of phase.
The slanted solid line marks $\psi_\perp$,
 i.e.~the net value of the X-mode PA calculated as 
a fixed-phase average of the bifurcated grey band. The dotted line marks the
RVM-based reference ($\psibe$).
The double-peaked microbeams, with arrows emerging from points A and B,
show how the PA is distributed within the neighbouring pulse longitudes. 
The void assumed in $\eta_\perp$ at D ($\phi=0$) creates the horizontal 
break in the bifurcated PA band (grey). 
The resulting imbalance of the PA averaging
produces the thick solid PA wiggle. 
}
\label{polidea}
\end{figure}

\begin{figure}
\includegraphics[width=0.48\textwidth]{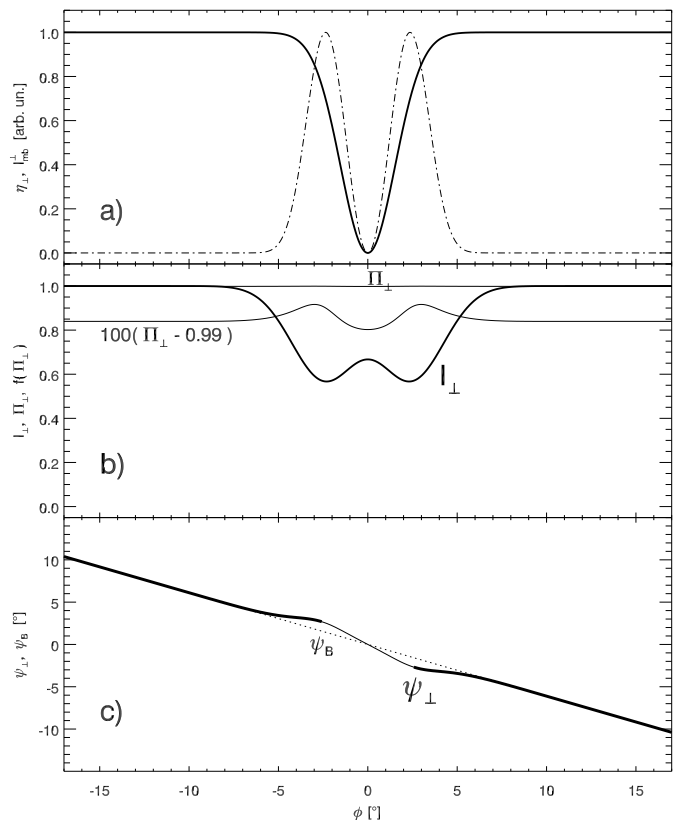}
\caption{Result of a calculation which produces 
the PA wiggle at  
a nearly merged double notch feature. {\bf a)} Microbeam intensity $\ibm$
(dot-dashed line) and the macroscopic emissivity with a Gaussian cavity at $\phi=0$
(thick solid).
{\bf b)}  Net polarisation fraction $\Pi_\perp$ (top thin solid)
and  net X-mode intensity profile (thick solid). The $\Pi_\perp$ is slightly
smaller than $1$, as shown by the bottom thin line presenting the quantity 
$100(\Pi_\perp-0.99)$.
{\bf c)}  Net X-mode PA ($\psi_\perp$, solid) 
overplotted on the reference
$\psibe$ (dotted). The central part of $\psi_\perp$ is thinner
to reflect the lack of high-quality data points at the centre of notches in 
B1821$-$24A (see Figs.~\ref{flux}d and \ref{lucas}d).
The result is for $\sigma_\eta=1.53^\circ$, $\psi_B=-0.6\phi$, $\rho_{\rm
crv}=5\times 10^4$ cm, and $\nu=1$ GHz.
}
\label{npol1}
\end{figure}

The convolution of the microbeam (dot-dashed line in Fig.~\ref{polidea}a) 
with the spatial emissivity $\eta_\perp$ (thick solid line)
 produces the double feature
shown by the thin solid line in Fig.~\ref{polidea}a.
This means that a microbeam such as the one shown in the bottom right
corner of Fig.~\ref{polidea}b (dotted line) is not contributing
its PA at the blank gap around point D 
in the greyscale distribution of PA.
The averaging of PA at the phase marked with C then leads 
to the upward deflection of the net PA because the flux at C is 
dominated by emission from the phase B, with a larger PA.
On the right-hand side of D, the same effect leads to a downward PA deflection
of the same magnitude.
The outcome is a zigzag-shaped wiggle of the net PA around the reference
$\psibe$.

Fig.~\ref{npol1} presents the calculation made for a Gaussian
void in the emissivity: $\eta_\perp = 1 - \exp(-0.5\phi^2/\sigma_\eta^2$),
with $\sigma_\eta = 1.53^\circ$ (thick solid line in panel a), 
and for the microbeam of the apparent 
half size $\dbm = 2.34^\circ$ (dot-dashed line).
The reference PA was assumed to change with phase
at a rate similar to that observed in PSR B1821$-$24A: 
$\psibe = -0.6\phi$ (dotted line in Fig.~\ref{npol1}c;
straight solid line in Fig.~\ref{flux}d). 
The net PA is shown in panel c 
by a thick solid line, which is made thinner in the central region
of the notches where the observed PA is unavailable because of the
insufficient S/N. 
There is 
%\LEt{ ok? the "one" construction should be avoided whenever possible } 
a flattening of PA in the outer wings of the notches,
which makes the impression of a discontinuous jump in PA similar to the
one observed in B1821$-$24A.
Panel b presents the intensity profile (thick solid line)
and the linear polarisation fraction $\Pi_\perp$ (top thin line).
It is clear  that the convolution of a single-mode (X-mode) radiation
with similar values of PA (as constrained by the double-peaked 
PA distribution
in Fig.~\ref{polidea}b)
does not produce perceivable depolarisation. A plot of $100(\Pi_\perp-0.99)$
reveals that outside the notches $\Pi_\perp$ drops by less than one per cent
and is larger at the minima because the contribution of depolarising
radiation (from a second lobe) is missing there.
The modelled change in PA at the notches (by about $6^\circ$) 
is two times smaller than observed in B1821$-$24A.

The increase in $\Pi_\perp$ and the small change in PA 
at the notches are
inconsistent with the observations of B1821$-$24A; however,
the result of Fig.~\ref{npol1} does not include the possible 
contribution of the other
polarisation mode, i.e.  the ordinary mode (O mode, or $\parallel$-mode), 
which is polarised parallel to the sky-projected B-field.
In the following we  use the term `net PA' to describe the 
net (average) PA of a single polarisation mode (either X or O).
For an average PA which contains
both modes we use the term `total net PA' (or just `total PA').
 
\subsection{Polarisation of notches in the presence of two modes}
\label{twomodes}

\begin{figure}
\includegraphics[width=0.48\textwidth]{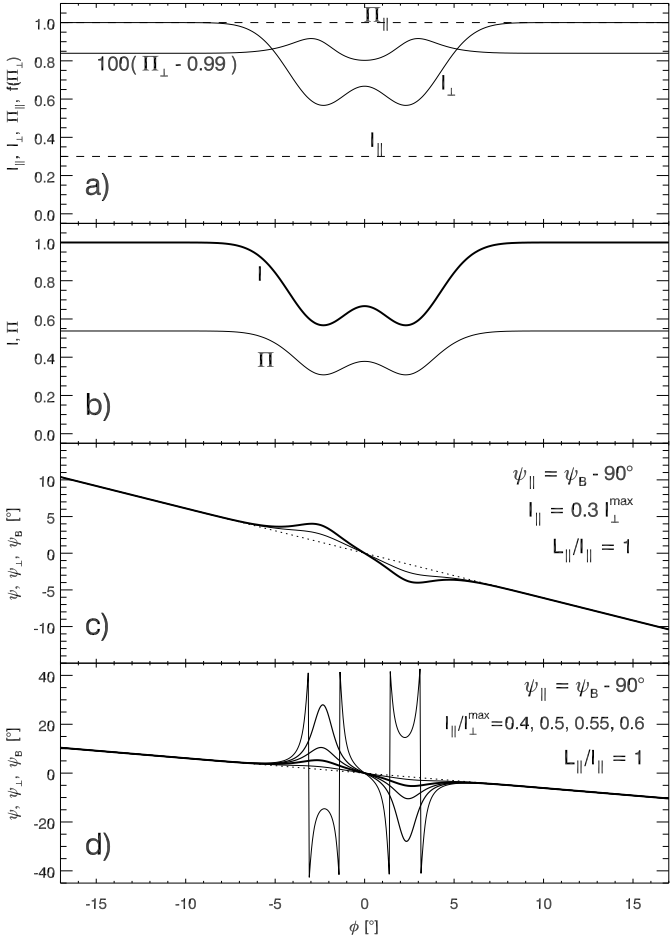}
\caption{Polarisation of double notches in the presence of
two polarisation modes. {\bf a)} Intensity and polarisation fraction
for both modes ($\perp$-mode: solid line; $\parallel$-mode: dashed line). 
$\Pi_\perp$ is indistinguishable
%indiscernible 
%\LEt{ indistinguishable? indiscernible = it can't be seen } 
from $\Pi_\parallel=1$, hence we plot $100(\Pi_\perp -
0.99)$. {\bf b)} Total intensity (thick solid line) and the total $\Pi$ (thin line).
{\bf c)}  Total PA $\psi$ (thick solid line) 
overplotted on the net PA of the X mode
 ($\psi_\perp$, thin solid line), and the reference $\psibe$ of the RVM model (dotted line). 
%We note 
%{\bf 
%Notice 
%} 
%how the
The contribution of the O mode increases the off-RVM 
amplitude of the total PA. {\bf d)} Same as in c, but for the increasing 
contribution of the O mode: $I_\parallel/I_\perp^{\rm max} 
= 0.4$, $0.5$, $0.55$, and $0.6$. The last case (thin line with spikes)
undergoes the orthogonal-mode jumps. 
The unspecified parameters are the same as in Fig.~\ref{npol1}.
}
\label{ipar}
\end{figure}

The effects of adding the O mode are illustrated in Fig.~\ref{ipar}, 
calculated
for the same parameters of the X mode as before. We have added a fixed
amount of the O mode ($I_\parallel = 0.3I_\perp^{\rm max}$) polarised
strictly at a right angle with respect to the reference (RVM) PA
of the X mode ($\psi_\parallel
= \psibe - 90^\circ$, i.e.~the O mode is polarised along the projected $B$
field). The linear polarisation properties of both modes are
shown in panel a, with $I_\perp$ referring to the net profile of
the X mode ($I_\perp$ results from the convolution 
of the curves shown in Fig.~\ref{npol1}a).
The total net profile (thick solid line in Fig.~\ref{ipar}b) 
shows a drop in $I$ down to $0.55$ at the notches' minima. The total
polarisation fraction 
$\Pi$ (thin line in panel b) is now decreasing from $0.54$  to $0.3$,
in  qualitative agreement with the observations of B1821$-$24A.
The wiggle of PA (panel c) extends over a larger interval because the O mode
`attracts' any non-orthogonal PA values to itself.
The presence of the other polarisation mode then tends to magnify
any slight deviations of the primary PA from  strict orthogonality.
Depending on the relative proportions of both modes, the amplitude of the
wiggle can easily be arbitrarily increased.
Fig.~\ref{ipar}d, calculated for $I_\parallel/I_\perp^{\rm max} 
= 0.4$, $0.5$, $0.55$, and $0.6$, presents how the PA deviation
($\Delta\psi=\psi-\psibe$) grows  with the increasing
contribution of the $\parallel$-mode. 
For the selected parameters (i.e.~for the assumed change in 
$\psibe$ across the microbeam width), 
$\Delta\psi$ exceedes $10^\circ$ for $I_\parallel$ approaching $I_\perp$
with an accuracy of $10\%$. 
The largest distortion 
(for $I_\parallel = 0.6I_\perp^{\rm max}$, thin line with spikes)
shows the jumps in  PA by $\pm90^\circ$ caused by the dominance of
the O mode over the primary X mode at the minima of the notches.

The ease of obtaining such large distortions 
is an important phenomenon which has a crucial role in shaping
the observed average PA curve: tiny distortions of PA from the reference
$\psibe$, become strongly magnified by the contribution of the other mode.
The original (small) distortions in a single mode naturally result
from the non-uniformity of the emission region (convolved with the
properties of the microbeam). This makes
the primary PA distribution at a given phase  a little asymmetric 
with respect to the reference $\psibe$. The incoherent contribution
 of the second polarisation mode can amplify 
these original PA deflections (characteristic of a single mode) 
to a very large magnitude, easily comparable to the $90^\circ$ separation
between the modes. When both polarisation modes have comparable flux,
 the net PA may be essentially arbitrary, which leads to
the randomisation of the fixed-phase PA distributions
described in McKinnon \& Stinebring (1998) and Melrose et al.~(2006). 
\nct{ms98}

At a fixed phase, each polarisation mode may then be expected to 
have the form of an intrinsic PA distribution with a finite width
and with a specific shape, which is in general 
asymmetric with respect to $\psibe$.
Therefore, the PA `curve' of pulsars should rather be understood as a
PA stripe or band, as is frequently viewed in the greyscale plots
presenting the PA distribution as a function of phase (e.g.~Stinebring et
al.~1984; Edwards \& Stappers 2004; Hankins \& Rankin 2010; 
Oslowski et al.~2014).  \nct{scr84, hr10, es04, hr10}
The total net pulsar polarisation can be understood as the interaction 
of these two quasi-perpendicularly polarised bands of PA. 
%\LEt{ just checking:  are the bands polarised almost perpendicularly?(perhaps: the interaction of these two bands of PA whose polarisations are almost perpendicular) or  are the   polarised bands   almost perpendicular to each other?  }
%{\bf [just checking:  are the bands polarised almost perpendicularly?
%(perhaps: the interaction of these two bands of PA whose polarisations 
%are almost perpendicular) 
%or  are the   polarised bands   almost perpendicular to each other?]
%The bands are NOT quasi-orthogonal to each other. 
%They are quasi-orthogonally polarised.
%The version suggested after "(perhaps:" is almost acceptable, though 
%I prefer the original (shorter) one.
%}
Any skewness of such a PA stripe (i.e.~of a single-mode PA histogram at a
fixed phase)
results in deflections which can be strongly enlarged by the second
polarisation mode. However, if both polarisation 
modes are observed simultaneously, only the total distribution of PA
may be perceivable at a fixed phase  and  the details of each mode
PA distribution undetectable. 
As explained in Section \ref{where}, even if the pure X mode is
detectable with negligible noise, the bifurcated PA distributions
are unlikely to be observable.
Our search for them
in J0437$-$4715 was indeed unsuccessful.

The agreement of Fig.~\ref{ipar} with the polarisation data 
on the double notches in B1821$-$24A is qualitative only.
For example, the modelled polarisation fraction $\Pi$ 
does not decrease to such a low value as observed, and is actually
slightly larger at the centre of the notches than at the minima.
The amplitude of the modelled PA swing barely reaches $10^\circ$
as compared to the observed value of $15^\circ$. 
There may be several reasons for these differences, 
and we discuss them in Sect.~\ref{summary}.

%including the simplicity
%of the model which assumes the perfect symmetry of 
%the effective microbeam and the fixed PA. 
%Other simplifications are the 
%uniform $\eta_\perp$ outside of the void, and the uniform $I_\parallel$
%everywhere.
%It is also likely that an extra excess of the O-mode appears at the center of
%notches in B1821$-$24A, because such a dramatic drop of $\Pi$ is not 
%observed in J0437$-$4715 (Fig.~\ref{monster}c). Such developments are certainly
%worth of implementing, however, they considerably enlarge the parameter
%space, which places the problem beyond the scope of this paper.
%Instead, it will be shown in the following, that the wiggle distortion
%of Fig.~\ref{npol1}c is generic enough, to be applicable for
%other pulsar polarisation phenomena, including the much %considerably 
%different behaviour of double notches in J0437$-$4715.

\begin{figure}
\includegraphics[width=0.48\textwidth]{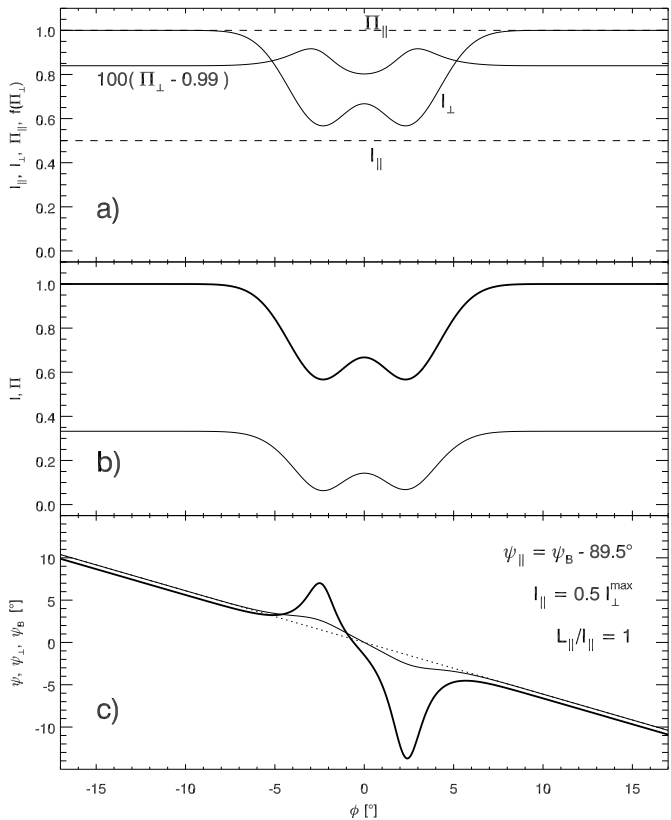}
\caption{Polarisation of double notches in the presence of 
a considerable amount of a quasi-orthogonal secondary polarisation mode
($\psi_\parallel = \psibe-89.5^\circ$, 
$I_\parallel= 0.5I_\perp^{\rm max}$). 
The layout is the same 
as in panels a-c of Fig.~\ref{ipar}.
%We note 
%{\bf 
%Notice
%}
%the large amplitude and asymmetry of the PA wiggle in panel c. 
The PA wiggle in panel c acquired large amplitude and became asymmetric.
Unspecified parameters are the same as before.
}
\label{npol2}
\end{figure}

\subsection{Interpreting the polarisation of PSR J0437$-$4715} 
\label{monstersec}

\begin{figure}
\includegraphics[width=0.48\textwidth]{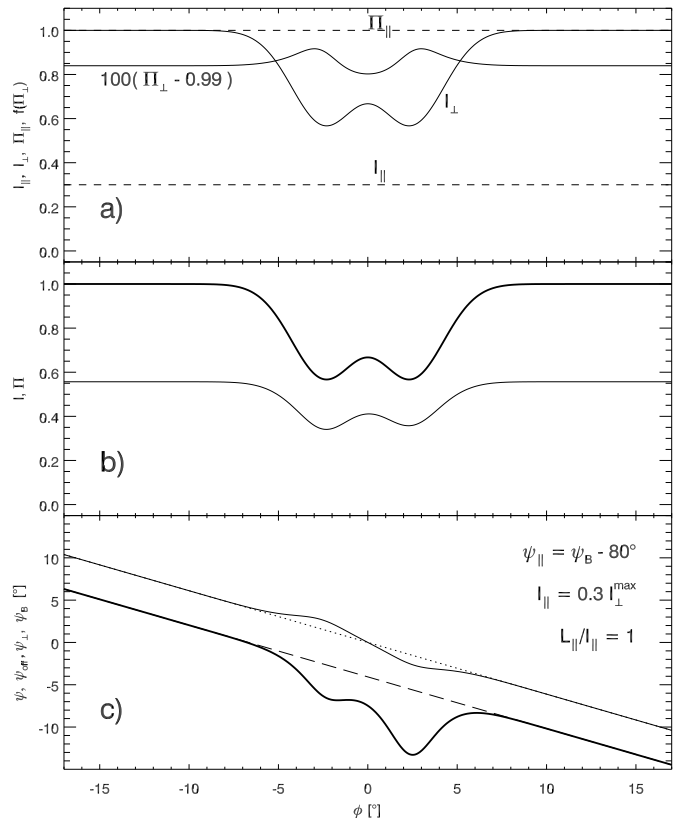}
\caption{Polarisation of double notches in the presence
of a slightly non-orthogonal secondary polarisation mode 
($\psi_\parallel = \psibe-80^\circ$, $I_\parallel= 0.3I_\perp^{\rm max}$).
The dashed line in {\bf c} presents $\psi_{\rm off}$, 
i.e.~the interpolated off-notch trend of the total PA. 
%We note 
%{\bf 
%Notice
%}
% that the 
The total PA at 
%\emph{
both
%} 
minima is located below $\psi_{\rm off}$
 (see the explanation in Fig.~\ref{stokes}).
}
\label{npol3}
\end{figure}

The secondary ($\parallel$) polarisation mode is subject to different
conditions of amplification and propagation (Melrose 2003; 
Wang, Wang \& Han
2015; Beskin \& Philippov 2012).  \nct{m03, wwh15, bp12}
It may undergo its own deflections from $\psi_B$, 
as determined by the skewness of O-mode PA distributions, 
resulting from the convolution of the O-mode emissivity 
with the O-mode microbeam. Since the mode is prone to additional propagation
effects, e.g.~a refraction, the PA of the O mode 
may possibly be even far from its RVM track.
In this paper all these O-mode deflections are parametrised
as a uniform shift of $\psi_\parallel$ with respect to $\Psi_B$,
where $\psi_\parallel$ denotes the net PA of the parallel mode.
We then assume $\psi_\parallel = \Psi_B+\epsilon_\parallel = \psibe-90^\circ
+\epsilon_\parallel$,
with a small or moderate value of $\epsilon_\parallel$. This simplification is done
to make interpretation easier because in real data the total 
net PA results from an interplay of the X-mode deflections 
with the deflections of the O mode. 

The summation of such quasi-orthogonal modes (Fig.~\ref{npol2}c)
slightly displaces the whole net PA curve towards $\psi_\parallel$,
increases the amplitude of the wiggle, and makes it asymmetric.
The wiggle of Fig.~\ref{npol2}c resembles those observed for 
J0437$-$4715 in the phase intervals:
$(-70^\circ,-58^\circ)$, $(-40^\circ,-14^\circ)$, 
and $(13^\circ,33^\circ)$ (indicated by rectangular boxes in 
Fig.~\ref{monster}b).
As in our simulation, the observed wiggles seem to be associated with 
minima in intensity, revealed by the abrupt changes 
in the intensity 
gradient, e.g.~at $\phi=-65^\circ$, $-10^\circ$, $10^\circ$, and $15^\circ$. 
They may also be traced to double minima in $\Pi$
 (grey line in Fig.~\ref{monster}a).
The wiggles may then be interpreted
as the result of the $\phi$-dependent skewness of the X-mode 
PA histogram, with the two-directional bias of PA 
induced by the broad absorption features. 
Since the observed width of these structures is several times
larger than the scale of double notches, their origin 
is likely dominated by the effects of macroscopic emissivity.

\begin{figure}
\includegraphics[width=0.48\textwidth]{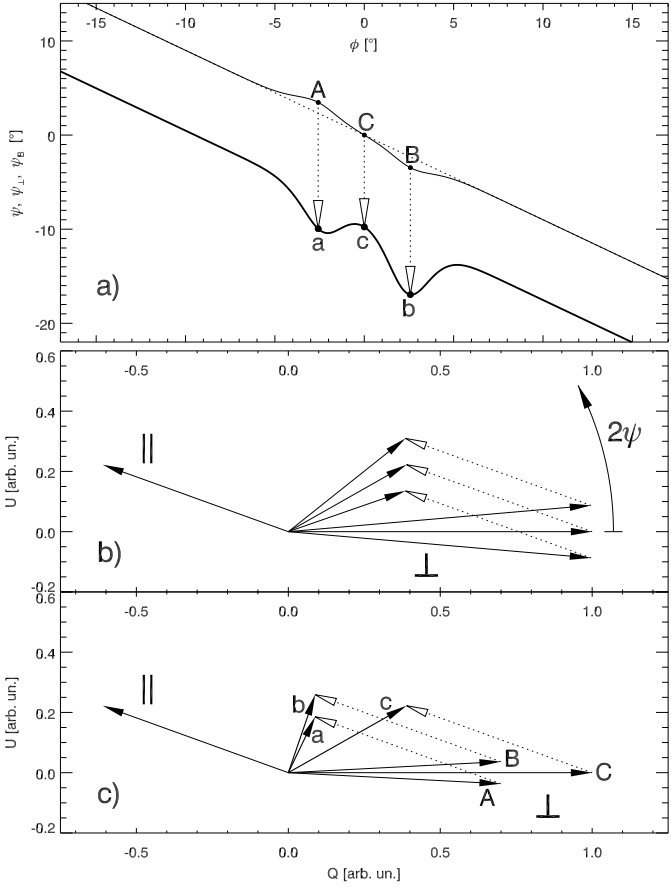}
\caption{ Mechanism of transformation of the bidirectional PA distortion
(the thin solid X-mode wiggle in panel a) 
into a one-directional W-shaped deflection of the total PA
(thick solid). The Stokes vectors associated with the marked points
are shown schematically in panel c. Those with small letters (a, b, c)
are the sum of the corresponding X-mode vectors (A, B, C)
with the quasi-orthogonal $\parallel$-mode vector shown on the left. 
The transition to the W-shaped PA is mainly caused by
 the drop in the X-mode flux at the minima of double notches 
(short vectors A and B).
Panel b presents a similar sum for the case with no drop in  polarised 
flux at A and B 
(all $\perp$-mode vectors have the same length).
}
\label{stokes}
\end{figure}

In the case of J0437$-$4715, the PA at 
both minima of the double notches deflects in the same direction,
which makes it different from B1821$-$24A.
This behaviour readily appears 
%\LEt{ is clearly seen? } 
%{\bf [is clearly seen?] No, the meaning of "readily appears" is: 
%"appears naturally, without
%fine tuning of parameters" so it may not be replaced with "is clearly seen".
%Perhaps: "naturally appears", though I prefer the original version.
%} 
in the model under some
circumstances: when the linearly polarised X-mode 
flux drops considerably at the minima and
the O mode is not too close to orthogonality 
with respect to the X mode (moderate $\epsilon_\parallel$).
In Fig.~\ref{npol3}c the O mode is $10^\circ$ off orthogonality,
so the overall PA curve is displaced down to the dashed line,
which marks the interpolated off-notch trend $\psi_{\rm
off}$ for the total PA.
Even at the leading-side minimum ($\phi\approx -2^\circ$),
where the net X-mode PA was above the RVM-based value ($\psi_\perp >
\psibe$),
the contribution of the O mode drags 
the PA across $\psibe$ and we get $\psi < \psibe$.
Remarkably, however, the resulting value of the total PA at that phase is 
even smaller than $\psi_{\rm off}$,
so the PA at both minima stays below the dashed line.

The value of $\psi$ at the centre of the notches does not undergo such a
large shift and remains closer to $\psi_{\rm off}$.
%the interpolated off-notch trend of the total PA.
The reason for which the PA at the leading-side minimum 
gets overdrawn to the other side of both $\psi_{\rm off}$ 
and $\psibe$ (as compared to its original
location) is illustrated in Fig.~\ref{stokes}. 
Because of the low polarised flux at the minima, the Stokes vectors 
representing $(L_\perp, \psi_\perp)$ (marked with A and B in
Fig.~\ref{stokes}c) are shorter than the vector C which corresponds
to the notches' centre. Therefore, the addition
of the same $\parallel$-mode vector (representing the fixed amount 
of the parallel mode)
is capable of inducing much larger rotation of PA towards $\psi_\parallel$
at the minima.
Without the drop in flux at the minima, this effect does not show up
(Fig.~\ref{stokes}b).

\begin{figure}
\includegraphics[width=0.48\textwidth]{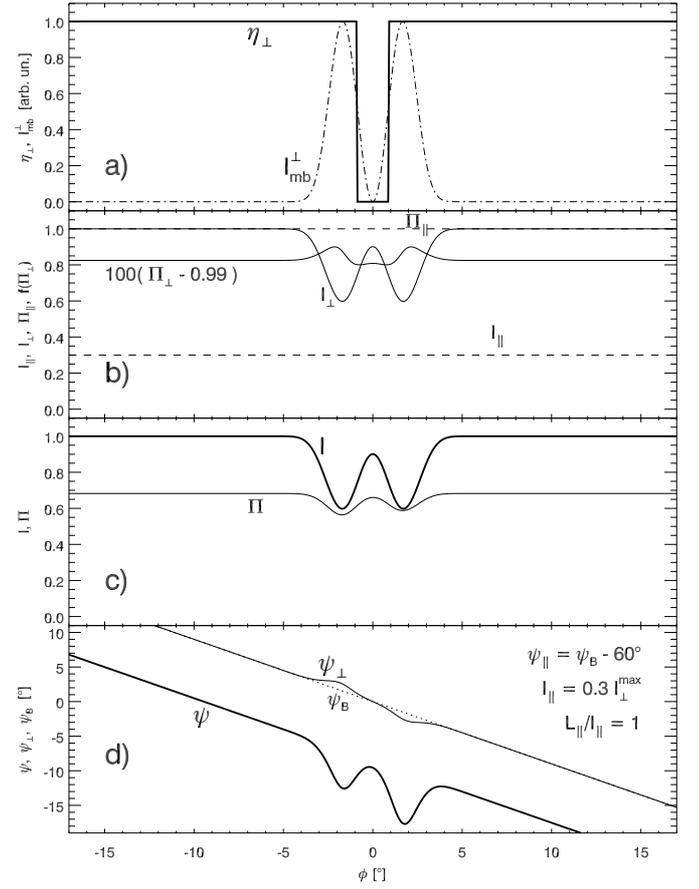}
\caption{Model results aimed at reproducing the observed
properties of the double notches in J0437$-$4715 (see Fig.~\ref{monster}).
A rectangular void in emissivity, of a half width equal to $0.9^\circ$
(thick line in a), has been assumed 
to raise the central maximum of I at $\phi=0$.  
The layout of panels b-d is analogous to panels a-c in Fig.~\ref{ipar}.
List of parameters:
$\psibe = -0.9\phi$, $\psi_\parallel=\psibe-60^\circ$, 
$I_\parallel=0.3I_\perp^{\rm max}$, $\rho_{\rm crv}=1.4\times 10^5$ cm, 
$\nu=1$ GHz. 
}
\label{narvoid}
\end{figure}

\begin{figure}
\includegraphics[width=0.48\textwidth]{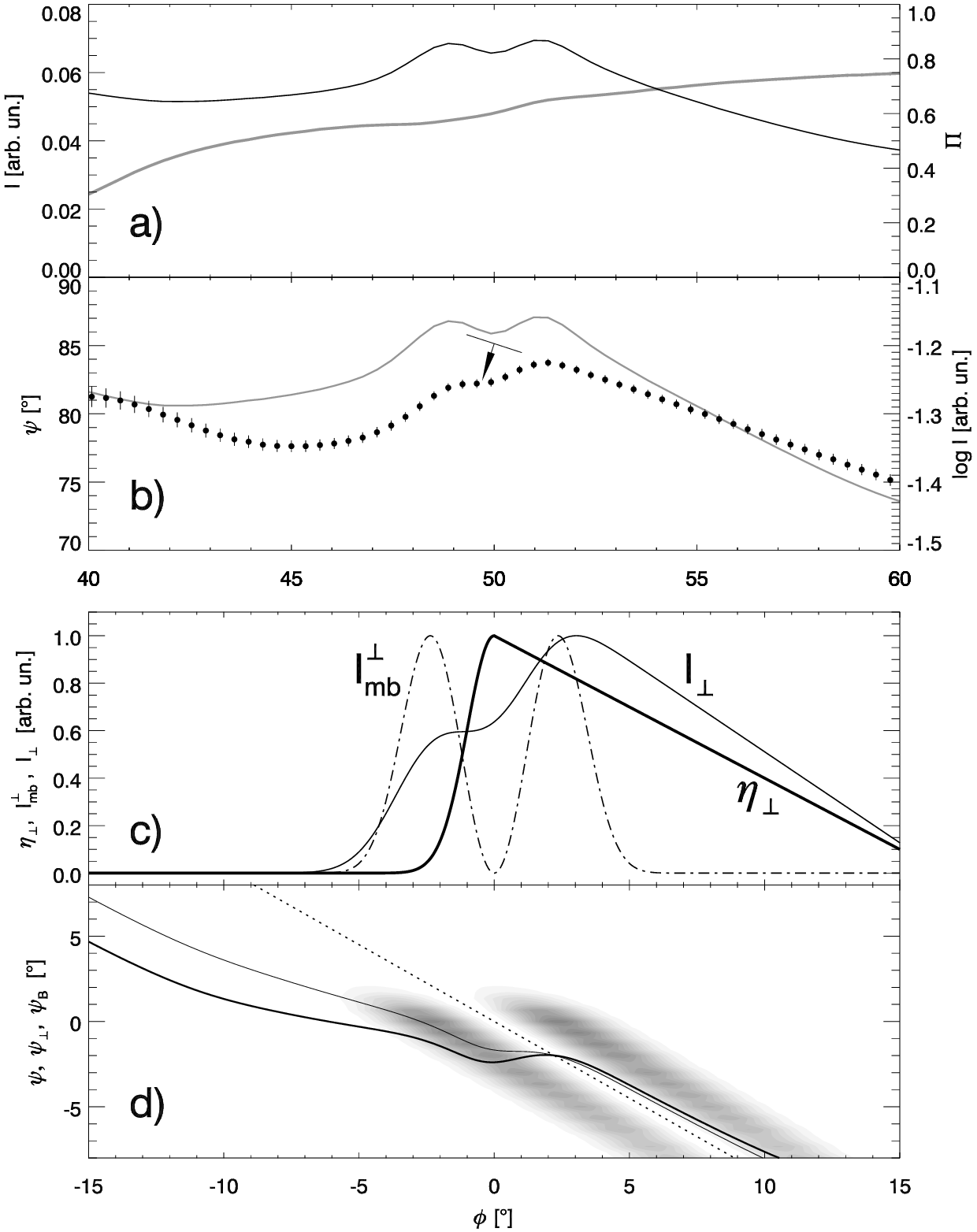}
\caption{Comparison of the observed (panels a and b)
and modelled (c and d) properties of the bifurcated trailing component (BTC) 
in J0437$-$4715. The observed properties ($I$ and $\Pi$ in {\bf a};
$\psi$ and $\log I$ in {\bf b}) present a zoomed part of Fig.~\ref{monster}.
The arrow in b marks a step-like change in PA at the centre of the BTC 
(the PA is shown with dots).
A similar step-like change in PA (solid lines in panel d) 
occurs for a one-sided emissivity profile ($\eta_\perp$, thick solid in c) 
which is decreasing gradually (linearly) 
on the right-hand side, but much more steeply on the left-hand side
(a half-Gaussian with $\sigma_\eta = 1^\circ$). 
The thin solid line in d presents $\psi_\perp$, whereas the thick line presents 
$\psi$
obtained for a fixed fractional contribution of the O mode: 
$I_\parallel=0.3I_\perp(\phi)$. 
This is an exemplary result
obtained for $\psibe=-0.9\phi$, $\rho_{\rm crv}=5\times 10^4$ cm.
}
\label{trsplit}
\end{figure}

Fig.~\ref{narvoid} presents a calculation performed
to roughly reproduce the characteristics of the notches in J0437$-$4715.
A rectangular void in $\eta_\perp$ has been assumed to reproduce the large depth
of notches with a high central maximum (the observed 
depth possibly approaches $\sim50\%$, though the absolute zero flux level
for J0437$-$4715 has not been determined rigorously).
For $\psibe = -0.9\phi$, $I_\parallel = 0.3I_\perp^{\rm max}$, and $\psi_\parallel=
\psibe-60^\circ$ the result roughly reproduces the $20\%$ drop in $\Pi$
and the $\sim\negthinspace\negthinspace5^\circ$ change in  PA 
(the observed deflections are 
$4^\circ$ at the leading-side minimum and $6^\circ$ at the trailing minimum).
The modelled deflection of PA reveals similar asymmetry.
However, the drop in $\Pi$ at the trailing minimum is a bit smaller
than at the leading minimum, in contrast with observations
(Fig.~\ref{monster}c). Possible reasons for these discrepancies are
discussed in Sect.~\ref{summary}.

% We have not attempted to model this, because the flux
%at the notches of J0437$-$4715 changes in a non-trivial way: the notches seem to 
%coincide with a broad emission bump overposed on a monotonically
%declining flux. For this reason it is difficult to precisely determine 
%several parameters, such as the height of the central maximum, or
%the fractional depth of the minima. The observed polarisation fraction $\Pi$
%does not change linearly around the notches, either.
%Qualitatively, however, the microbeam model with two polarisation modes 
%is capable of explaining why the PA behaviour 
%at the notches of J0437$-$4715 is so different from that of B1821$-$24A.

%trsplit

The observed PA exhibits an interesting behaviour at the bifurcated 
trailing component (BTC) in J0437$-$4715 (Fig.~\ref{trsplit}).
On the right-hand side of the BTC, 
the PA appears to change linearly; however, it undergoes a step-like drop 
at the centre of the BTC (dots in panel b).
Such a step-like change in PA is naturally expected in our `fixed-PA
microbeam model' if the emissivity $\eta_\perp$ changes more steeply
on the leading side of the BTC. 
Fig.~\ref{trsplit} (c and d) presents a sample result
obtained for $\eta_\perp$ which  
rises quickly on the BTC's left-hand side (following a Gaussian), but
decreases linearly on the right
 (thick line in c).
The step-like drop in PA occurs because the PA makes a transition 
from the single-lobe-dominated value on the left-hand side
to the net value close to $\psibe$ on the right-hand side 
 (see the thin 
solid curve for $\psi_\perp$ in panel d).
 If a fixed fraction of the O mode is added
($I_\parallel=0.3I_\perp(\phi)$, $\Pi\approx0.54$), 
then the total PA assumes the shape shown with the thick solid line in panel d.
The model result does not reproduce the relative flux of the BTC's peaks
(see the thin solid line for $I_\perp$ in panel c).
However, the calculation does not take into account 
several factors that influence the ratio.
These include 1) the likely asymmetry of the flux and PA in the effective
microbeam, which appears for 
a non-orthogonal traverse of the sightline through a non-uniform
split-fan beam (see Fig.~2 in DR12);
2) the profile of the O mode,
the amount of which is increasing towards the left-hand side of the BTC, 
as suggested by the observed $\Pi$ (grey line in a); and 3)
the relative steepness of $\eta_\perp$ 
on both sides of the BTC (a guess form 
%\LEt{ an estimated form? } 
%{\bf [an estimated form?] No, $\eta_\perp$ was essentially guessed, 
%not estimated. Please use the original version.
%Actually this issue depends on the definition of the word "estimate".
%A dictionary offers two different meanings: 
%1) a valuing or rating by the mind, without actually measuring;
%2) rough or approximate calculation.
%The second meaning is standard in science, which is  
%is why "guess form" is better, since no calculations, even of
%order-of-magnitude style, were performed.
%} 
of $\eta_\perp$ 
was used in Fig.~\ref{trsplit}). 
Detailed modelling of this feature
is deferred to a future study.

\subsection{Non-bifurcated microbeam}

\begin{figure}
\includegraphics[width=0.48\textwidth]{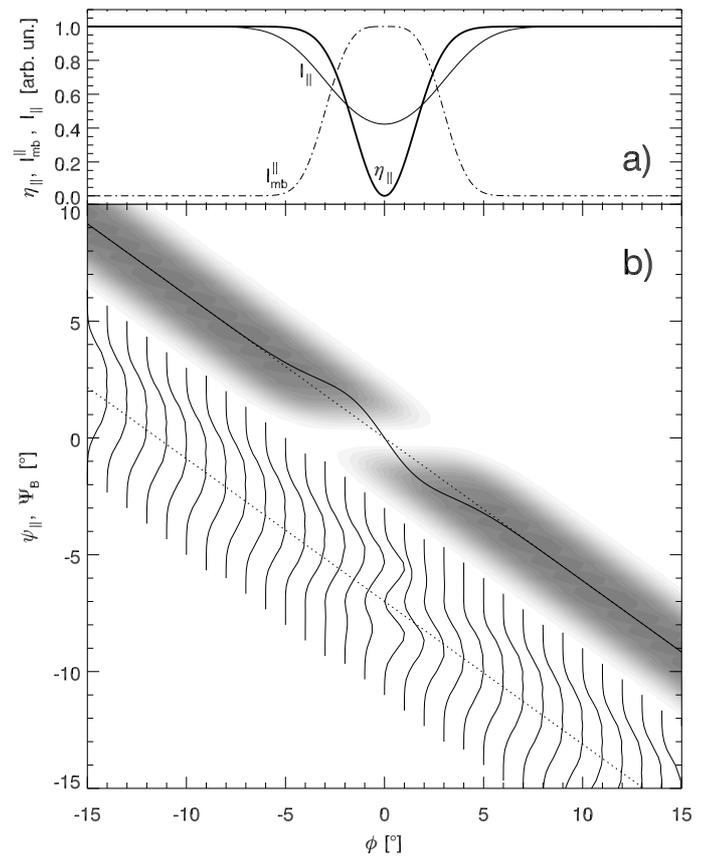}
\caption{ Origin of the zigzag-shaped PA wiggle in the case of
 the filled-in microbeam of the O mode. The layout is similar to that of
Fig.~\ref{polidea}, except for the fixed-phase PA distributions
which are plotted at $1^\circ$ phase intervals in panel b.
The minimum in $\eta_\parallel$ produces
 the single-minimum dip in the net O-mode intensity 
($I_\parallel$, thin solid line in a) as well as the
blank gap in the non-bifurcated grey PA band (panel b). 
The PA wiggle is again produced by the imbalance of the PA averaging.
}
\label{single}
\end{figure}

 In the case of the centripetal acceleration
the ordinary mode is emitted into a single-peak microbeam
 of non-negligible width,
such as that shown with the dot-dashed line in Fig.~\ref{single}.
We assume that the perceived PA is the same
at any direction within the beam and equal to the projected
direction of the $B$-field ($\Psi_B$).
In this subsection the void in the emissivity will be applied for the  O mode
only, $\eta_\parallel$ refers to the macroscopic emissivity of the O mode,
 and $\ibmpar$
represents the O-mode microbeam pattern.
By analogy to the  X mode (eq.~\ref{iper}), it is assumed that 
the O-mode microbeam 
 can be approximated by that part of the vaccum CR beam which is polarised
in the plane of the guiding $B$-field line, 
i.e.~$\ibmpar\propto \xi^2 K^2_{2/3}(y)$ (Rybicki \& Lightman 1979). 
\nct{rl79}

%\emph{
If the other polarisation mode is absent, 
%}
 then a small-amplitude 
wiggle of PA also appears for the non-bifurcated O-mode beam.
This occurs because the single-mode deflections depend 
on the asymmetry (or skewness)
of the PA distributions at a fixed-phase. 
In the case of the single-peaked microbeam,
the asymmetry is produced by the outer
wings of its radiation pattern (Fig.~\ref{single}).
 Because of the beam's extension,
a narrow dip in emissivity (thick solid line in panel a)
results in the horizontal break in the grey PA 
distribution shown in panel b.
The resulting fixed-$\phi$ histograms of PA, which are plotted 
every $1^\circ$ below the grey PA band, 
are clearly skewed in opposite directions on either  side of $\phi=0^\circ$.
The net PA then exhibits a wiggle similar to the one for the bifurcated
beam, albeit of a smaller amplitude. 
Close to $\phi=0^\circ$, where the flux is low,
the PA distribution becomes double 
(the fixed-phase PA histograms 
in Fig.~\ref{single}b
are normalised to the same peak value).

\begin{figure}
\includegraphics[width=0.48\textwidth]{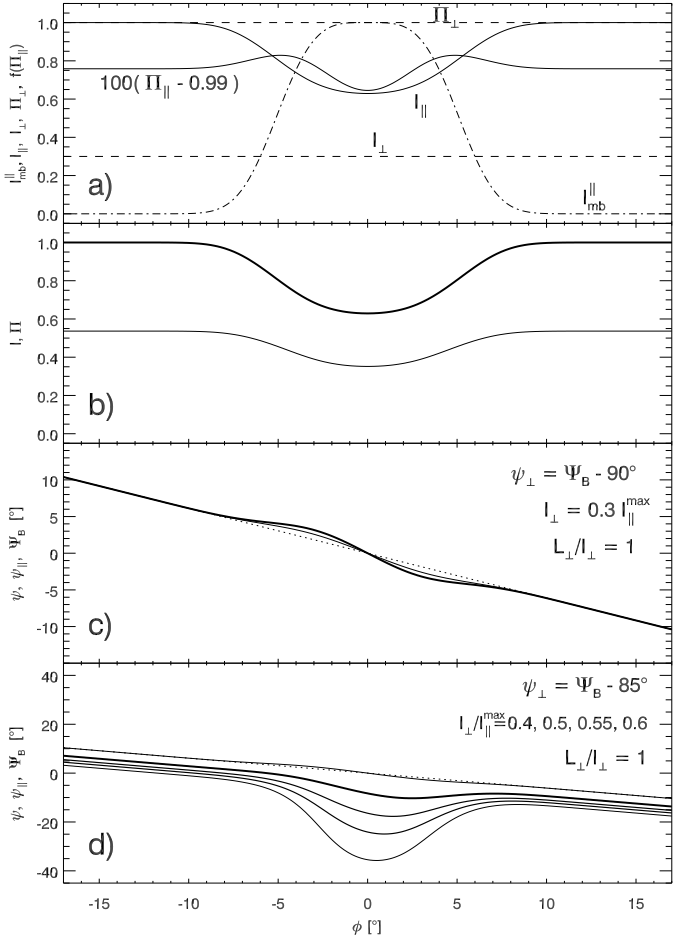}
\caption{ Influence of the secondary polarisation mode
 (this time it is the X mode) on the polarisation
of a single notch dominated by the primary O mode. 
A Gaussian void 
in emissivity $\eta_\parallel$ of width $\sigma_\eta=2.53^\circ$ 
is now applied only for the O-mode emission. 
The layout is analogous to that of Fig.~\ref{ipar}, except
in panel d the modes are not precisely orthogonal 
($\psi_\perp=\Psi_B-85^\circ$). 
%We note 
%{\bf
%Notice 
%}
%that the
The 
contribution of the X mode does not much increase the amplitude 
of the PA wiggle in panel c (cf.~Fig.~\ref{ipar}c). The slight 
non-orthogonality of $5^\circ$
(panel d) transforms the zigzag-shaped wiggle into a one-directional
U-shaped distortion of PA (imagine a variant of Fig.~\ref{stokes}c
with the vector C shorter than A and B). List of parameters:
$\psi_\parallel=-0.6\phi$, $\rho_{\rm crv}=10^4$ cm, $\nu=1$ GHz.
}
\label{iparsin}
\end{figure}

In the presence of both modes, however, the total PA has different
properties than in the previously described case of the void 
convolved with the bifurcated microbeam.
%{\bf 
This occurs because 
%(((The merged sentence is extremely long and complicated.
%PLEASE use the split version.)))
%}
 the void in $\eta_\parallel$ produces
a single dip in $I_\parallel$, with the minimum value
 ($I_\parallel=I_\parallel^{\rm min}$) 
at the centre ($\phi=\phi_{\min}=0$), whereas the maximum deflection 
of PA occurs in 
the wings of the feature (at $\phi_{\rm max} \ne 
\phi_{\rm min}$).\footnote{The subscript `min' refers
to the minimum flux of the primary polarisation mode (here $I_\parallel$),
whereas `max' to the maximum deflection of this primary mode PA from
the reference PA (here $\Delta\psi_\parallel=\psi_\parallel-\Psi_B$).
If the void is in $\eta_\parallel$, we
have a single notch with $\phi_{\rm min} \ne \phi_{\rm max}$.
For a narrow void in $\eta_\perp$, we have  double notches with 
$\phi_{\rm min} \approx \phi_{\rm max}$.} 
Therefore,
when the contribution of the other mode (X)  is increasing, 
$I_\perp$ first reaches $I_\parallel$ at $\phi_{\rm min}$ 
and produces the
orthogonal PA jump there before it is able to considerably distort 
$\psi$ at $\phi_{\rm max}$.  
If the modes are strictly orthogonal outside the notches 
($\psi_\perp = \psibe$
and $\psi_\parallel = \Psi_B$), 
a large-amplitude wiggle similiar to those shown 
in Fig.~\ref{ipar}d can only appear for finely tuned  
parameters:
$L_\perp \approx L_\parallel(\phi_{\rm min})$ (and $L_\perp < L_\parallel$) 
with accuracy of the order of $1$\%.
Otherwise, the  
%considerable contribution of radiation with
%$\psi\approx\Psi_B$, emitted in the center of the microbeam, 
%does not allow 
wiggle's amplitude does not increase considerably or the PA jumps by
 $90^\circ$ at $\phi_{\rm min}$.
For  modes that are not exactly orthogonal ($\psi_\perp \ne \psibe$),
the bidirectional deflection of PA %(thin solid in Fig.~\ref{ipar}c,d) 
(i.e.~the wiggle of Fig.~\ref{iparsin}c)
quickly acquires a U-shaped form (Fig.~\ref{iparsin}d), 
i.e.~it becomes a feature with a single minimum 
instead of either the zigzag shape or the  W shape.
This occurs  because 
%\LEt{ `This is because' is not acceptable stylistically } 
the total PA is mostly determined by
the U-shaped drop in the net O-mode flux %near $\phi=0$ 
(see the curve for $I_\parallel$ ($\approx\negthinspace L_\parallel$)
%shown as the thin line 
in panel a) 
and much less by the small zigzag deflection 
of the net PA (panel c). 
As shown in Fig.~\ref{stokes}c, 
%\emph{
the total PA is mostly determined
by the amount of the linearly polarised flux of the primary mode,
 rather than its PA.
%} 
This flux ($L_\parallel$) is minimal at the centre of
the U-shaped feature, hence the largest deflection of the total PA
occurs there (Fig.~\ref{iparsin}d). Thus, the total PA also follows
a U-shaped curve, just like the $I_\parallel$ (or $L_\parallel$) does.

We  then conclude that the lack of emission at the centre of the
microbeam facilitates the appearance of the wiggle-shaped
 deflections of PA in the average profiles.
This is consistent with the fact that the profiles of
PSR J0437$-$4715 and B1821$-$24A contain both the wiggle-shaped PA deflections
and the bifurcated features (the notches and the bifurcated
emission components).

\subsection{PA deflections at profile emission components}
\label{emicom}

\begin{figure}
\includegraphics[width=0.48\textwidth]{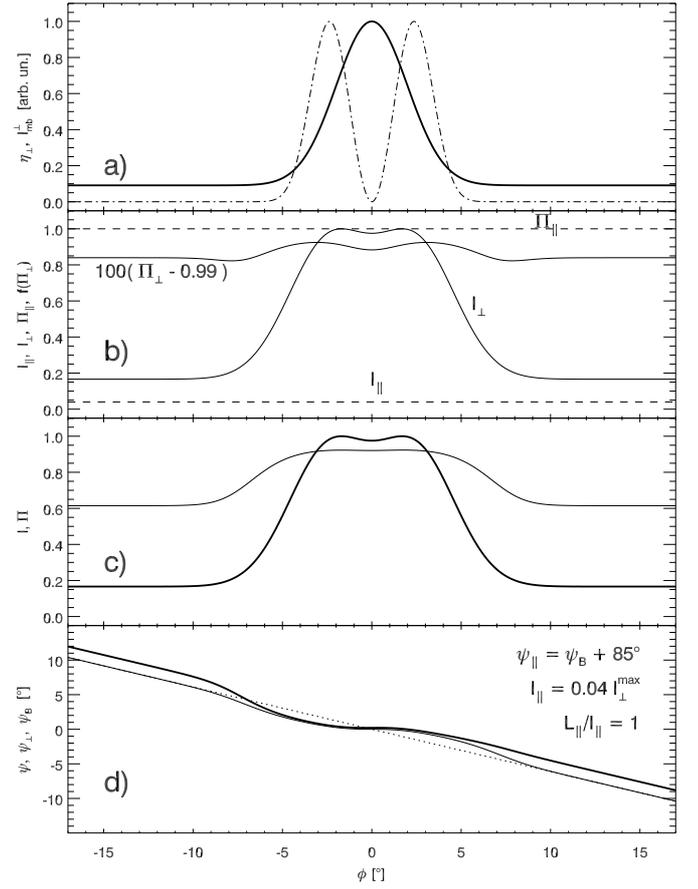}
\caption{Modelled polarisation of a bright narrow component projecting from
an isotropic background emission. 
%We note 
%{\bf 
%Notice 
%}
The PA curve exhibits 
 a flattening  
under the
component (panel d), caused by the fixed PA attributed to the microbeam.
A small uniform contribution of the O mode ($I_\parallel = 0.04
I_\perp^{\rm max}$) produces the off-centred, one-directional 
distortion of PA (thick line in d). 
Used parameters: $\psi_\parallel = \psibe+85^\circ$,
 $\rho_{\rm crv}=5\times 10^4$ cm.
%$\psi_\parallel=-0.6\phi$, $\rho_{\rm crv}=5\times 10^4$ cm, $\nu=1$ GHz.
}
\label{emi}
\end{figure}

In the case of PSR B1821$-$24A, some small distortions of PA appear 
under its two
brightest components (P1 and P2), both of which are double. 
The wiggle of PA at the second brightest peak 
(P1 at $\phi=-105^\circ$ in Fig.~\ref{flux})
looks deceptively similar to the modelled distortion 
described for the absorption features
(Figs.~\ref{npol1}c, and \ref{ipar}c,d).
On the other hand, the brightest peak in the profile of B1821$-$24A 
(P2 at $\phi=0$ in Fig.~\ref{flux}), has the PA which deflects in one direction
only (no zigzag) and the deflection is not phase-aligned with the
peak;  instead, it is located on its leading side.

%They may also be caused by the convolution 
%of the bifurcated microbeams, since the effects described for the
%void in emissivity, also occur for a narrow peak in emissivity.
%The case of emissivity peak is just a negative version
%of the void case, so all the effects described for the single void case, 
%must also emerge for the single spike in $\eta_\perp$.

As shown in Fig.~\ref{emi}, the fixed-PA microbeam model can easily produce 
off-centred one-directional
PA distortions, 
%\emph{
albeit not of the type 
%} 
observed for the peaks of B1821$-$24A.
Fig.~\ref{emi} has been
 calculated for a slightly wider maximum in the X-mode emissivity, 
$\eta_\perp = 0.1 + 0.9\exp(-0.5(\phi/2^\circ)^2)$, to reproduce the nearly
unresolved main component of B1821$-$24A (P2). When
a small amount (only $0.04I_\perp^{\rm max}$) 
of a quasi-perpendicular O mode is added ($\psi_\parallel=
\psibe+85^\circ$), the linear polarisation fraction
stays very high within the component (panel c) and the total PA
makes a single deflection on its leading wing (thick line in d).
The zigzag disappears because the convolution of the two modes 
aligns the off-notch trend of the total PA  with the net
X-mode PA in the trailing half of the component.

This result is somewhat similar to the observed one; however, 
unlike in the observation, 
the PA in Fig.~\ref{emi} is 
%\emph{
decreasing 
%}
with increasing pulse phase $\phi$, i.e.~the PA gradient $S=d\psibe/d\phi$ is
negative. Had we changed the sign of $S$ to the observed (positive) value, 
the one-directional deflection would
move to the trailing side of the peak, in conflict with the observation.
Had we additionally changed the sign of the modal non-orthogonality 
(taking $\psi_\parallel=\psibe-85^\circ$), the total 
PA deflection would move to the leading side, but it would be protruding
upward, again  inconsistent with data.
The reason is that the addition of a bright X-mode 
component to a low-level X-mode background 
always results in a flattening of the net PA gradient
under the added component (thin solid line in Fig.~\ref{emi}d);  whenever the microbeam width is larger than (or comparable
to) the width of the peak in $\eta_\perp$, the added component 
contributes the fixed PA value, so a
brightening in a profile should always be associated with the flattening
of the PA curve.
In contrast with this implication, 
the PSR B1821$-$24A (though not J0437$-$4715) 
exhibits  increased $|S|$ 
under the bright emission components.  
A possible reason for this discrepancy may involve unrecognised 
single-pulse effects (cf.~Melrose et al.~2006). 

%, in comparison 
%to the value of $|S|$ observed in phase intervals adjacent to the components. 
  
%This disagreement may be caused by the non-uniform (time-dependent) 
%emissivity of individual charges which move in the streams 
%that produce the components P1 and P2.
%The present version of our simple model 
%assumes the fixed-PA in the entire effective microbeam. 
%The change of emissivity near the line of sight
% results in the imbalance of the polarised intensity averaging, 
%which has the consequence that 
%the effective PA undergoes a fast change within 
%a phase interval comparable to the observed microbeam scale $\Delta$. 
%This is because the microscopically-local PA directions within the microbeam
%(see Fig.~1 in CR79) become detectable. \nct{cr79}
%Depending on the angular gradient of the emissivity 
% near the line of sight, 
%these PA variations can easily become faster than RVM. 
%In general, then, the shape and magnitude of PA deflections from RVM
%depend on the relative scale of the microbeam as compared to the angular
%scale of macroscopic emissivity gradients in the emission region.
%In this paper we took into
%account just the lateral gradients (in the direction perpendicular 
%to the $B$-field line planes). The fixed-PA assumption is equivalent to
%the uniform emissivity of individual charges, when they pass 
%through the sightline 
%in their longitudinal motion (along the stream). A full 3D study
%of polarisation will be presented elsewhere.

An observational detail that may be of great importance
is that a single orthogonal-mode jump 
(at $\phi \approx 70^\circ$ in Figs.~\ref{flux}b and \ref{lucas}b) 
separates  the P2 
from the double notches in B1821$-$24A. Therefore, if the notches are
interpreted as the signature of the bifurcated X-mode beam, 
the highly-polarised 
P2 (and also P1, with its PA being an extrapolation of the linear trend
observed near P2) should be dominated by the O-mode emission in the form of the 
filled-in (non-bifurcated) beam. The bifurcations of the P1 and P2 should
then  probably be interpreted in terms of macroscopic properties of the
emitter or in terms of  propagation effects. 
%\LEt{ yes? }  
A possible origin of such macroscopic bifurcations has been suggested in
Dyks \& Rudak (2015, see figure 12 therein).

\begin{figure}
\includegraphics[width=0.48\textwidth]{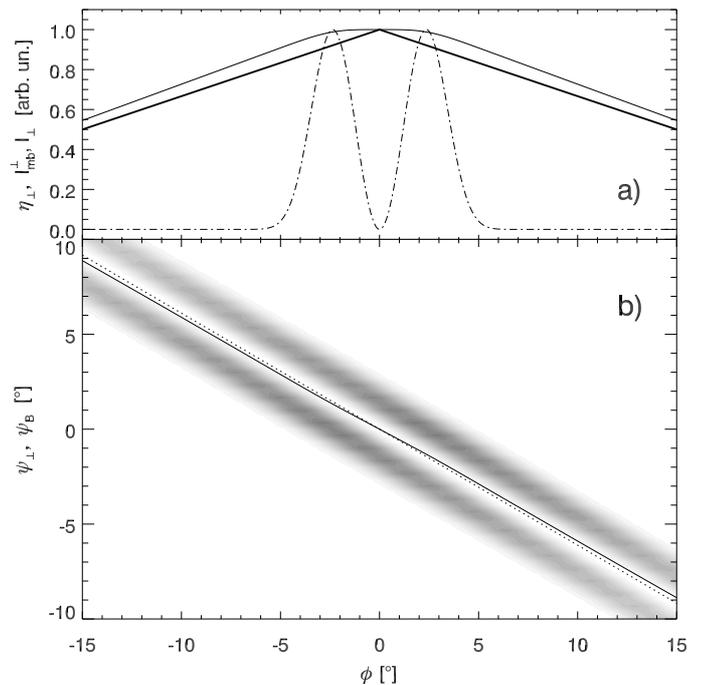}
\caption{Polarisation of a very wide emission structure with the triangular
emissivity pattern shown in panel a (thick solid line). 
%We note 
%{\bf 
%Notice
%}
%that the 
The net PA (solid line in panel b) nearly
 follows $\psibe$, but not exactly.
}
\label{emibump}
\end{figure}

Our model of polarisation 
(based on a wide microbeam) 
implies that any changes in flux in the profile
must be unavoidably accompanied by small deviations of PA
from the RVM.
When the profile emission components extend for wide intervals of phase,
and when  the flux changes slowly, then the PA distortions are not 
visible as clear local features;
however, the net PA is not likely to precisely 
follow the RVM-based 
value. This is shown in Fig.~\ref{emibump} where the X-mode $\eta_\perp$ 
has the form of a wide triangle (thick solid in panel a). 
The gradual distortion of the net PA (flattening near the flux maximum) 
is visible in panel b.

\subsection{Double-peaked PA distributions} 
%\subsection{Where are the double-peaked PA distributions}
\label{where}

The presented model is supported by the existence of the bifurcated
features in the profiles of B1821$-$24A
and J0437$-$4715. 
 Superficially, however, one might expect 
a stronger support to come from a direct
observation of the double-peaked PA distributions, such as shown in
grey in Fig.~\ref{polidea}b. 
 The peaks of the pure X-mode PA distributions are separated by
$\Delta |d\psi_B/d\phi|$, where $\Delta$ is the observed scale of double
features. In most cases discussed before $|d\psi_B/d\phi|\sim1$, 
thus the separation is of the order of a few degrees.
We have searched for the PA bifurcations in J0437$-$4715 
 using the polarisation calibrated single-pulse data described in detail in 
Os{\l}owski et al.~(2014). \nct{ovb14}
%a single-pulse data from the Parkes telescope.} 
The search was unsuccessful,
 even within this part of the profile which contains
bifurcated features.

This can be understood, however, and does not pose a problem 
for our model because 
the flux detected at any moment is
a convolution of several intrinsic signals (not to mention
the instrumental noise) which can strongly affect the observed 
two-dimensional
$(\phi, \psi)$ distributions. 
Specifically, the flux in each peak of the double 
PA distribution (as observed at a fixed phase) corresponds to emission from two
different (but nearby) phases, and two different (but closely located) 
emission points in the magnetosphere. For example,
at the phase C in Fig.~\ref{polidea}, one of the PA peaks
originates from phase B, and another one  from phase D.
%In real-time detection (single-pulse observations), 
%the PA peaks may then have very different flux and look single.
%On the other hand, 
If the two emitting points
permanently contribute both these PA peaks,
only their average (close to $\psi_B$) will be detectable at any moment 
in single-pulse observations.
Therefore, for the PA bifurcation to be visible even in the pure X mode, 
non-simultaneous contributions of 
%comparable-strength 
flux from both phases
should occur. During one star rotation, the flux at the considered phase 
would have to be
strongly dominated by emission from one phase (e.g. B), 
during another rotation  by the other phase (D).
The bifurcation will be lost if the contributions 
are comparable and simultaneous.
Since we discuss
two nearby pulse phases and a laterally extended emission region,
such temporal separation of emission seems unlikely, which explains
why the PA bifurcations are not observed.

Thus, to see the bifurcation, the flux which contributes to one PA peak
has to be received in different pulses (or at least in different 
%or adjacent
%{\bf ((("or adjacent" may be removed here)))}
samples) than the flux which contributes to the other PA peak.
%(and which corresponds to different star rotations or samples). 
The PA values in both peaks may therefore
be affected by the instrumental noise of different sign or strength.
This additionally increases the difficulty of detection of the PA
bifurcation, even when the conditions of the previous paragraph are
fulfilled. 

The detection may even be made more difficult 
by the simultaneous emission in both
polarisation modes. In this case we observe the instantaneous 
total PA determined by the degree of their non-orthogonality (and by
the relative amounts of both modes).

We then conclude that the clear bifurcation
of the X-mode PA, such as illustrated in Fig.~\ref{polidea}b,
is not likely  to be detected unless several `purifying'
conditions are met. The main condition requires that the emission in
each PA peak occurs in the above-described successive way, 
belonging either to the first or to the second PA peak.
This is further complicated by the real-time instrumental noise,
and the possible contribution of the O mode.
Our model is then consistent with 
the ubiquitous lack of the PA bifurcations in pulsar data.

\subsection{Multiple PA bands in pulsars}

The issue of simultaneous emission of both polarisation modes
is related to the possibility of observation of several non-orthogonal 
PA stripes on the $(\phi,\psi)$ plane.
This phenomenon is reported, for example,  
%{\bf  e.g. 
%(((what is wrong with this "e.g." here? I wanted to inform that
%B1237$+$25 is likely not the only object that exhibits this feature)))
%}
 for B1237$+$25 observed at 327 MHz
at the Arecibo Observatory (Smith et al.~2013). \nct{srm13}
A distinct PA stripe, which looks like a wiggle which is not parallel 
to the PA stripe of the primary mode, 
is observed under the central component 
in the profile of this object. If the primary mode is assumed to follow
the RVM, the distinct additional wiggle
may be interpreted as the PA stripe corresponding
to the simultaneous emission of both modes. The primary PA stripe
would then represent the undisturbed PA of the primary mode 
emitted mostly alone.

In the simplest case there are three distinct situations in which 
only the primary or secondary mode is emitted, or both modes are emitted 
simultaneously. 
Then at least three PA stripes can be produced
on the $(\phi, \psi)$ plane, one corresponding to the primary mode,
another to the secondary mode (if it happens to be emitted alone
%\LEt{ yes? emitted needs an adverb (alone); solitary is an adjective }
), 
and a third stripe for the simultaneous emission of both modes.
The location of the last (non-RVM) stripe is determined
by the intrinsic degree of non-orthogonality 
and the relative flux in both modes. As shown in Melrose et
al.~(2006), for a well-defined (sharp) 
extra PA stripe,
the  radiation emitted simultanously   must be characterised 
by a preferred (frequently occurring) combination of these parameters. 
Otherwise the PA becomes randomised (McKinnon and Stinebring 1998).
\nct{ms98}
Examples of such randomisation can sometimes be found in 
single-pulse PA data; see~e.g.~the profile of B1919$+$21 
in Hankins \& Rankin (2010) and  \nct{hr10}.
%see van Straten \& Tiburzi (in preparation) for a more thorough
%analysis of the effects of mode mixing.
%{\bf 
Thorough analysis of mode mixing effects is presented in van Straten \& Tiburzi 
(in preparation).
%(((if possible, I would like to avoid the merging of the last sentence, 
%so I separated and rephrased it)))
%}

%Another explanation involves the microbeam scale
%and/or the longitudinal emissivity gradient of the stream which produces the core
%component. The rest of the profile (not the core) may be 
%produced by microbeams of negligible size in comparison to the scale of 
%emissivity gradients in the emission region. Therefore, the PA follows
% the RVM under  most of the profile. In the case of the core component, 
%however, the microbeam scale may be comparable to the macroscopic scale
%(e.g.~the emissivity may drop steeply along the core stream, or the microbeam
%may be wider, because of a smaller curvature radius of electron trajectories,
%or both). Then the core PA would tend to partially follow the
%``microbeam PA swing", which is steeper than RVM (see Section \ref{emicom}). 
%Depending on whether 
%the strong core is present or not, single pulse observations would record
%the steep microscopic swing, or the RVM swing, respectively.

%An alternative explanation for the non-RVM
%character of the wiggle would invoke propagation effects, which 
%operate in one mode only and make it deflected from the RVM.

\section{Discussion}
\label{summary}

%Pulsar profiles for which the PA can be measured within extended
%intervals of pulse longitude, exhibit local distortions of PA from the
%overall PA curve. These are usually associated with localised emission
%or absorption features, some of which are bifurcated (double).
%Detailed properties of the distortions cover a wide range 
%of behaviour, including the zigzag-shaped, W-shaped, or U-shaped 
%deflections of PA, associated with large or small decrease of polarisation
%fraction, and with a steepening or flattening of the PA gradient.

To account for observed PA distortions,  a CR-based 
model can be set up which
assumes emission into a bifurcated beam of non-negligible angular extent 
and   polarised at a fixed angle with respect to the local $B$-field. 
This elementary beam (a microbeam) is convolved with an 
emissivity profile which represents the lateral extent of the emission
region, and it is supplemented with a contribution 
of the secondary polarisation mode. 

A model of such a type is capable of 
qualitatively reproducing the polarisation 
behaviour of several dissimilar features in the profiles of 
B1821$-$24A and J0437$-$4715. 
This provides additional support for the stream-shaped geometry 
of the pulsar emission region, and for the fan-shaped geometry
of the pulsar beams (Michel 1987; DRD10; DR12; 
Wang et al.~2014; Dyks \& Rudak 2015). 
\nct{m87, dr15, wpz14, dr15}
It also shows that CR is a useful mechanism for interpreting the
 pulsar radio emission.

%However, these peaks
%are likely dominated by the O-mode radiation, so their polarisation
%may be determined by different effects than assumed for the X mode.
%To interpret this behaviour with the fixed-PA microbeam, the microbeam
%needs to be manipulated, e.g.~by postulating partial obscuration 
%in plasma cavities.

%In spite of the partial success in the reproduction of observed
%polarisation effects, the model fails to account for a few details.
%In addition to  
However, the model cannot reproduce
the steepening of PA under emission components, nor 
 the large drop in the polarisation 
degree at the centre of notches in B1821$-$24.
The notches may be created by a single non-emitting 
plasma stream embedded in a laterally extended emission region, 
or as a dense obscuring stream, located above the emission region
 (see fig.~12 in DRD10). If the bifurcation of the notches 
originates from the
extraordinary-mode nature of the microbeam, 
then the non-emissive interpretation
is favoured because the X-mode radiation does not interact with the
plasma in the superstrong $B$-field; 
hence, it should not be absorbed or obscured. 
It is possible that the `dark' stream is responsible for the
giant pulses observed near the centre of the notches (Figs.~1 and 2 in
BPDR). The stream may either be directly generating the giant 
pulse emission, or it may be reprocessing the background radiation of the
surrounding emitter, e.g.~via scattering of the radio photons.
In either case
the new type of radiation may have different polarisation properties
from the surrounding background which could produce the extra
depolarisation. 
%We note 
%{\bf 
%Notice 
%} 
It is important to note that
such a contribution of extra emission
at the centre of the double notches can mislead our modelling 
efforts
since the height of the central maximum of double notches 
is the main factor which
determines the width of the void in emissivity $\eta_\perp$.

 The model predicts that the 
%\emph{
intrinsic 
%} 
bifurcation of the PA
distribution at a fixed phase is unlikely to be detectable, unless 
%\emph{ 
special 
%} 
conditions are satisfied. 
These include not only the nearly pure X-mode emission
 (mostly free from the noise and the O mode), but also a sequential
(non-simultaneous) detection of flux from two nearby pulse longitudes 
(those which contribute each PA peak at a considered pulse phase).
The proposed model is therefore consistent with the lack
of the PA bifurcations in the pulsar data.

Generally, the model illustrates the key importance of the 
circum-RVM distribution of PA for the shape of the average PA curve.
Since the spread of PA at a fixed phase is intrinsic, it is worth
 thinking in terms of a PA band or stripe instead of a PA curve.
 These fixed-phase PA distributions are unlikely to be symmetrical
around the RVM-based value, and any small deviations of one mode 
are amplified by the existence of the other polarisation mode.
This can lead to  large PA deflections in the averaged profiles, 
and to the PA randomisation in single pulses. 

This work suggests that in several pulsars 
the angular scale of the microbeam is not negligible in comparison 
to the angular gradients of emissivity in the emission region.
Many observed distortions from the RVM may result from 
 the microscopic PA becoming recognizable despite the convolution
with the spatial extent of the emitter. 

The data reproduction achieved through the by-eye
fitting described in Sect.~\ref{results} is qualitative only, and 
the main reason for this are the guessed 
%\LEt{ estimated? `guess' sounds very arbitrary }
%{\bf [estimated? `guess' sounds very arbitrary] 
%Please use "guessed", not "estimated".
%} 
and simple forms of $\eta_\perp$ 
and $\eta_\parallel$.
Instead, the
observed polarised intensity profiles likely do not correspond to 
any simple analytical functions. For example, the notches in J0437$-$4715
seem to coincide with a broad emission bump superposed on a monotonically
declining flux, whereas the BTC consists of $\eta_\perp$ and $\eta_\parallel$,
which cannot be described by a simple exponential or linear function.
A possible solution to this problem would be to split the observed average
profile into two fully polarised orthogonal modes, then deconvolve 
the microbeam from the mode-separated profiles to learn the approximate form
of $\eta_\perp$ and $\eta_\parallel$. Another development that may be
needed is a three-dimensional code capable of simulating an oblique 
traverse through the fan beams (hence including the apparent asymmetry 
of the microbeam). Moreover, the single-pulse population effects
 (Melrose et al.~2006) have already been shown to crucially affect
the apparent polarisation. They seem to be the most important ingredient 
that may need to be included in the modelling to achieve close
agreement with the data. 
%This would increase the number 
%of free parameters, so
%the assistance of additional observational
%constraints (eg.~distributions of polarised samples at various pulse
%longitudes) 
%would be required.
Thus, the question of whether complicated 
average PA curves (such as that of J0437$-$4715) can be disentangled
into their pure RVM form, remains open.

\section*{Acknowledgements}
JD thanks A. Bilous and S. Ransom for 
providing us with the profile of B1821$-$24A before publication.
 SO is supported by the Alexander von Humboldt Foundation.
The Nan\c{c}ay Radio Observatory is operated by the Paris Observatory,
associated 
with the French Centre National de la Recherche Scientifique (CNRS). 
This work was supported by 
the National Science Centre grant DEC-2011/02/A/ST9/00256.
\bibliographystyle{/home/jinx/MARCO/PSEP/Bib_Ref_Style/aa}
%\bibliographystyle{mn2e}
%%\bibliography{listofrefs,/home/jinx/MARCO/PSEP/Bib_Ref_Style/journals,/home/jinx/MARCO/PSEP/Bib_Ref_Style/psrrefs,/home/jinx/MARCO/PSEP/Bib_Ref_Style/crossrefs,/home/jinx/MARCO/PSEP/Bib_Ref_Style/modrefs,/home/jinx/MARCO/PSEP/Bib_Ref_Style/pierba}

\bibliography{listofrefs}

%\appendix

%\section{The conal $R_W$-distribution in the flat geometry}
%\label{flatderiv}

%When the beam is narrow ($\rout \ll 1$ rad) and the dipole tilt is large
%($\alpha \gg \rout$) the $R_W$ distribution is well approximated by the

\end{document}